\documentclass[12pt,a4paper]{article}
\usepackage{amsmath}
\usepackage[font={footnotesize,it}]{caption}
\usepackage[top=1in, bottom=1in, left=1in, right=1in]{geometry}
\DeclareCaptionStyle{italic}[justification=centering]{labelfont={bf},textfont={it},labelsep=colon}
\usepackage[title]{appendix}
 
\usepackage{graphicx,psfrag,epsf,textcomp}
\usepackage{lineno}
\usepackage{setspace,authblk}
\usepackage{enumerate}
\usepackage[round]{natbib}
\usepackage{url,xcolor} 
\usepackage{booktabs, bm, paralist,mathpazo,tikz,longtable,microtype}
\usepackage{xcolor,soul,amsfonts,setspace,mathrsfs,mathtools,placeins}
\usepackage{multirow,array}
\usepackage{float}
\usepackage{subcaption}
\usepackage{longtable}

\usepackage[pdftex,colorlinks=true]{hyperref}
\definecolor{darkblue}{rgb}{0,0,.6}
\definecolor{a0}{rgb}{0.0, 0.5, 0.0}
\definecolor{bistre}{rgb}{0.24, 0.17, 0.12}
\definecolor{amethyst}{rgb}{0.6, 0.4, 0.8}
\definecolor{Rcolor}{RGB}{150,160,190}
\definecolor{blush}{rgb}{0.87, 0.36, 0.51}
\definecolor{brightturquoise}{rgb}{0.03, 0.91, 0.87}
\definecolor{burntorange}{rgb}{0.8, 0.33, 0.0}
\hypersetup{citecolor=darkblue,linkcolor=darkblue,urlcolor=darkblue}
\newcommand*\patchAmsMathEnvironmentForLineno[1]{%
  \expandafter\let\csname old#1\expandafter\endcsname\csname #1\endcsname
  \expandafter\let\csname oldend#1\expandafter\endcsname\csname end#1\endcsname
  \renewenvironment{#1}%
     {\linenomath\csname old#1\endcsname}%
     {\csname oldend#1\endcsname\endlinenomath}}%
\newcommand*\patchBothAmsMathEnvironmentsForLineno[1]{%
  \patchAmsMathEnvironmentForLineno{#1}%
  \patchAmsMathEnvironmentForLineno{#1*}}%

\AtBeginDocument{%
\patchBothAmsMathEnvironmentsForLineno{equation}%
\patchBothAmsMathEnvironmentsForLineno{align}
\patchBothAmsMathEnvironmentsForLineno{flalign}%
\patchBothAmsMathEnvironmentsForLineno{alignat}%
\patchBothAmsMathEnvironmentsForLineno{gather}%
\patchBothAmsMathEnvironmentsForLineno{multline}%
}

\newcommand{\blind}{1}

\graphicspath{{images/}}

\newcolumntype{L}[1]{>{\raggedright\let\newline\\\arraybackslash\hspace{0pt}}m{#1}}
\newcolumntype{C}[1]{>{\centering\let\newline\\\arraybackslash\hspace{0pt}}m{#1}}
\newcolumntype{R}[1]{>{\raggedleft\let\newline\\\arraybackslash\hspace{0pt}}m{#1}}

\newsavebox\CBox

\date{}
\usepackage{orcidlink}

\begin{document}

\def\spacingset#1{\renewcommand{\baselinestretch}
{#1}\small\normalsize} \spacingset{1}

\if1\blind
{
\title{\bf A Compositional Approach to Modelling Cause-specific Mortality with Zero Counts}
\author[1]{\normalsize Zhe Michelle Dong}
\author[2]{\normalsize \quad Han Lin Shang}
\author[1]{\normalsize \quad Francis Hui} 
\author[1]{\normalsize \quad  Aaron Bruhn}
\affil[1]{\normalsize Research School of Finance, Actuarial Studies and Statistics, The Australian National University, Canberra, Australia}
\affil[2]{\normalsize Department of Actuarial Studies and Business Analytics, Macquarie University, Sydney, Australia}
\date{}
} 

\maketitle
\fi

\begin{abstract}
Understanding and forecasting mortality by cause is an essential branch of actuarial science, with wide-ranging implications for decision-makers in public policy and industry. To accurately capture trends in cause-specific mortality, it is critical to consider dependencies between causes of death and produce forecasts by age and cause coherent with aggregate mortality forecasts. One way to achieve these aims is to model cause-specific deaths using compositional data analysis (CODA), treating the density of deaths by age and cause as a set of dependent, non-negative values that sum to one. A major drawback of standard CODA methods is the challenge of zero values, which frequently occur in cause-of-death mortality modelling. Thus, we propose using a compositional power transformation, the $\alpha$-transformation, to model cause-specific life-table death counts. The $\alpha$-transformation offers a statistically rigorous approach to handling zero value subgroups in CODA compared to \emph{ad-hoc} techniques: adding an arbitrarily small amount. We illustrate the $\alpha$-transformation on England and Wales, and US death counts by cause from the Human Cause-of-Death database, for cardiovascular-related causes of death. Results demonstrate the $\alpha$-transformation improves forecast accuracy of cause-specific life-table death counts compared with log-ratio-based CODA transformations. The forecasts suggest declines in proportions of deaths from major cardiovascular causes (myocardial infarction and other ischemic heart diseases (IHD)). 

\vspace{.1in}

\end{abstract}

\spacingset{1.5}

\section{Introduction}\label{sec:introduction}

Understanding mortality by cause is key to informing medical research decisions and planning social services \citep{KEK+19, AAS15}. It is also important in assessing mortality rates and longevity risk for life insurers, as causal factors can drive the best estimate of mortality and morbidity assumptions for the purposes of reserving and pricing. Analysing and modelling cause-of-death data presents two main challenges: the need to account for inherent dependencies between various causes of death and the need to produce forecasts by causes coherent with aggregate mortality forecasts. 

Traditional methods for mortality modelling and forecasting, including the Lee-Carter (LC) model \citep{LC92} or variations thereof and the age-period-cohort model \citep{H83, RH06}, among others, generally do not account for dependencies between competing causes of death. As such, over the past two decades, considerable progress has been made on joint models for multiple causes of death, which capture between-cause dependencies. \citet{AS13} applied vector error correction models (VECM) to cause-of-death mortality rates to quantify the dependence between competing risks and subsequently found an improvement in forecasts compared to methods that do not allow for such dependencies. \citet{AAS15} formulated a multinomial logistic model across several causes of death to investigate the effects of improvement and elimination of mortality due to cancer. \citet{LLL19} adopted a forecast reconciliation approach to ensure coherence in cause-specific mortality rates, while \citet{LL19} introduced hierarchical Archimedean copulas to capture dependence between competing risks in causes of death. More recently, \citet{ZHH23} developed a predictive approach for cause-of-death mortality modelling that jointly models various causes, ages, and years using a penalised tensor decomposition. 

The majority of literature on modelling mortality by cause, including those mentioned above, treats cause-specific life-table deaths as non-compositional, that is, through modelling age-specific mortality rates rather than age distributions of death. Although these methods enable modelling dependencies between mortality rates for different causes, a more direct approach is to forecast the cause-specific death distribution, where the dependence is explicitly incorporated by capturing relativities between deaths of one cause and another. Indeed, cause-of-death data are fundamentally compositional, as deaths have been recorded and attributed to various causes for analysis in globally used medical classifications for epidemiology, health management, and understanding mortality experience \citep{ICD10}. 

With this in mind, an alternative approach, known as compositional data analysis (CODA), has arisen in the actuarial science literature, which aims to model the cause-specific death distribution directly and produce mortality forecasts arising from the composition of the distribution itself. The idea of CODA dates back to the seminal work of \citet{A82} for analysing data that arises as a vector of observations where the elements sum to a constant value and, therefore, only contain relative information. In the context of mortality by cause, the compositional sum constraint translates to avoided deaths from one cause, leading to increased deaths from other causes. 

When analysing compositional data, methods that ignore the compositional constraint and apply standard multivariate data analysis to the raw observations (we refer to such approaches as ``raw data analysis" or RDA) can encounter potential issues with coherence when it comes to aggregated mortality forecasts. An alternative approach in CODA is to transform the compositional data from the simplex, subject to the unit sum constraint, to the unconstrained real space before applying standard multivariate data analysis and forecasting. Then, the results are transformed into the compositional space for interpretation and inference. Within this latter approach, log-ratio transformations are by far the most widely used to transform compositional data due to their various attractive compositional properties \citep[see][for details]{A82}. 
The first to propose such a ``log-ratio analysis" or LRA for forecasting mortality rates was \citet{O08}, who applied an LC mortality model to log-ratio transformed death compositions to forecast cause-specific mortality. \citet{O08} used centred log-ratio transformation (see Section~\ref{sec:modelling} for details) and found that capturing dependencies between subgroups via LRA and the CODA framework improved the overall forecast while assuming independence between causes tended to produce pessimistic results, that is, expected deaths tend to be overstated. \citet{KEK+19} further extended this approach by developing two new LRA models for cause-specific deaths, adding cause-specific weights to age and time subgroups, and decomposing joint and individual variation between causes of death to improve forecast accuracy further. Other notable works include that of \citet{BCO17}, who applied CODA to produce age-coherent forecasts for mortality, \citet{BSS+22}, who used LRA to model healthy life expectancy, and \citet{ KEB+20}, who produced longevity forecasts by socio-economic group using LRA.

Whilst the aforementioned works use LRA to address some of the issues with analysing compositional data (relative to RDA), one outstanding challenge with LRA-based modelling is the presence of zero counts/values \citep{BCO17, KEK+19, KEB+20}. Specifically, compositional data with zero values can be interpreted as lying on a boundary of the simplex. So, naively applying a log-ratio transformation to such data results in one or more transformed values taking $\pm \infty$. In the context of mortality by cause, zero death counts in subcategories of the composition arise commonly for new and emerging or granular causes of death at certain ages and at older ages where exposure is limited. Since the existence and treatment of zeros may lead to differences in the overall inference and forecasts, as mentioned above, this could have consequences on our understanding of longevity risk and mortality improvements, along with associated financial implications \citep{B13}. 

In the literature, the problem of zeros when using LRA has often been addressed in an \emph{ad-hoc} manner by omitting, aggregating, or adding small arbitrary values to zero values \citep{MBP03}. For instance, \citet{KEK+19} explored imputing half of the minimum observed death count, a method initially used by \citet{BCO17}. Alternatively, \citet{KEK+19} noted that \citet{HBY13} imputed death rates based on information from nearby years for the same age group using linear interpolation. None of these methods is ideal; furthermore, \citet{G21} compared four different algorithms to substitute zeros and showed the resulting conclusions could be susceptible to the technique of zero substitution. More recently, \citet{G23} introduced the $\chi$-power transformation to address the problem of zeros in compositional data by combining the chi-squared distance in correspondence analysis with the Box-Cox power transformation. 

In this article, we propose a novel approach to modelling mortality by cause with zero values using a modification of LRA. We introduce a compositional power transformation known as the $\alpha$-transformation \citep{TPW11}, which addresses the challenges presented by zero values in the setting of CODA in a more statistically principled manner compared to the aforementioned \emph{ad-hoc} techniques. The $\alpha$-transformation, which maps compositional data to remove their unit sum constraint, is a generalised Box-Cox power transformation that includes both RDA and LRA as special cases but more broadly involves a tuning parameter $\alpha \in (0,1]$. This parameter can be calibrated in a data-driven manner to enable more flexibility in producing forecasts compared to standard LRA when there are zero values in the data. While the $\alpha$-transformation has been applied to CODA for geology and biology, amongst other fields \citep{TS20}, to our knowledge, this paper is the first to examine its use in forecasting mortality by age and cause.

We apply the $\alpha$-transformation to two datasets: 16 years of cause-of-death data from England and Wales data, and 43 years of cause-of-death data from the US. In both applications, we disaggregate for cardiovascular causes such that there are data with zero counts in one or more subgroups.
We couple the $\alpha$-transformation with the LC mortality model for multivariate analysis and forecasting \citep[similar to those of][]{O08, KEK+19}, and compare results with several LRA and RDA approaches where \emph{ad-hoc} methods are used to deal with zero values. Results across both applications demonstrate the $\alpha$-transformation generally improves mortality forecast by cause, while having the added benefit of being able to analyse compositional data with zero counts in a rigorous yet data-driven manner. The $\alpha$-transformation is shown to address the key issue of zero counts in mortality data, generalising the log-ratio transformation to a broader class of transformations and providing additional flexibility and improved performance when forecasting mortality by cause using CODA-based techniques.

The remainder of this paper is structured as follows: Section~\ref{sec:modelling} reviews several key ideas, including the Lee-Carter (LC) mortality model and log-ratio analysis. Section~\ref{sec:alpha transformation} introduces the $\alpha$-transformation for mortality by cause data. Section~\ref{sec:application} applies the proposed methodology to forecast mortality on cause-of-death data from England and Wales and the US, while Section~\ref{sec:discussion} offers some concluding remarks.

\section{Review of Key Concepts}\label{sec:modelling}

We review three foundational concepts for understanding how the $\alpha$-transformation can be applied to cause-of-death mortality modelling, namely compositional data (Section~\ref{sec:Compositional Data}), log-ratio analysis or LRA (Section~\ref{sec:log-ratio transformations}), and the LC mortality model (Section~\ref{sec:LC-CODA}).

\subsection{Compositional data}\label{sec:Compositional Data}

Cause-specific mortality can be represented by actual death counts per combination of year, age group, and cause. Specifically, let $D_{t, u, c}$ denote the actual death count for year $t = 1, 2, \dots, T$, age group $u = 1, 2, \dots, U$, and cause $c = 1, 2, \dots, C$, and define $D_t = ~\sum_{u=1}^U \sum_{c=1}^C D_{t, u, c}$ as the total deaths across all age bands and cause groups for year $t$. Then we can calculate $d_{t, u, c} = D_{t, u, c}/D_{t}$ such that for a given year, the vector $\bm{d}_t = (d_{t,1,1},d_{t,1,2},\ldots,d_{t,1,C},d_{t,2,1},d_{t,2,2}, \ldots, d_{t,2,C}, d_{t,u,1},d_{t,u,2}, \ldots,d_{t,U,C})$ represents the density distribution of deaths by age group and cause. The densities in $\bm{d}_t$ are ordered such that the cause runs faster than age. Moreover, the compositional vector satisfies $\sum_{u=1}^U \sum_{c=1}^C d_{t, u, c} = 1$. Moreover, by stacking the $\bm{d}_t$'s as row vectors on top of each other, we can form the $T \times UC$ compositional matrix $\mathbf{D}$ of death densities
\begin{equation}\label{eqn:D Matrix}
    \mathbf{D} = \begin{pmatrix}
        d_{1, 1, 1} & d_{1, 1, 2} & \dots & d_{1, 1, C} & d_{1, 2, 1} & d_{1, 2, 2} & \dots & d_{1, U, C} \\
        d_{2, 1, 1} & d_{2, 1, 2} & \dots & d_{2, 1, C} & d_{2, 2, 1} & d_{2, 2, 2} & \dots & d_{2, U, C} \\
        \vdots & \vdots & \ddots & \vdots & \vdots & \vdots & \ddots & \vdots \\
        d_{T, 1, 1} & d_{T, 1, 2} & \dots & d_{T, 1, C} & d_{T, 2, 1} & d_{T, 2, 2} & \dots & d_{T, U, C} \\
    \end{pmatrix}.
\end{equation}
Due to the sum-to-one constraint, only $UC-1$ elements are needed to uniquely determine each vector $\bm{d}_t$. Statistically then, the sample space for compositional cause-of-death mortality data is a simplex: for all $t = 1, \dots T$,
\begin{align*} 
    S^{UC-1} = \left\{ (d_{t, 1, 1}, \dots, d_{t, U, C}) | d_{t, u, c} \geq 0,\quad \sum_{u = 1}^{U} \sum_{c = 1}^{C} d_{t, u, c} = 1 \right\}.
\end{align*}

\subsection{Log-ratio analysis}\label{sec:log-ratio transformations}

A common approach to analysing compositional data is to employ the log-ratio transformations class, which seeks to transform the data from the simplex back to an unconstrained real space before building a statistical model for analysis. The two most common types of transformations within LRA are the centred log-ratio (CLR) and isometric log-ratio (ILR) transformations, which we consider in this paper. Importantly, the CLR and ILR are used for analysing compositional data \emph{without} zero values.

The CLR transformation is defined by dividing all the values in the compositional vector by their geometric mean before applying the natural log transformation. For row $t$ in~\eqref{eqn:D Matrix}, the CLR for each element is given by
\begin{align}\label{eqn:CLR}
    w(d_{t, u, c}) &= \ln\left(\frac{d_{t, u, c}}{(\prod_{u=1}^U \prod_{c=1}^C d_{t, u, c})^{1/UC}}\right) = \ln(d_{t,u,c}) - \frac{1}{UC} \sum_{u=1}^U\sum_{c=1}^C \ln(d_{t,u,c}).
\end{align}
The CLR transformation is symmetric relative to the compositional parts and has the same number of components as the number of parts in the original composition. We can express the CLR-transformed vector as $\bm{w}(\bm{d}_t) = (w(d_{t,1,1}),w(d_{t,1,2}),\ldots,w(d_{t,U,C}))$, noting distances between any two elements of this vector remain the same when measured in the simplex and the real space, thus making the CLR particularly useful for analysis \citep{GOE19}. While each element is no longer constrained to be non-negative (in principle, they can take any real number), the entire vector remains constrained since the elements must sum to zero by the construction of~\eqref{eqn:CLR}.

To further remove this constraint, the ILR left matrix multiplies the CLR transformed vector by a Helmert sub-matrix and has been promoted as the more theoretically correct method (especially to contrast groups of elements) in CODA \citep{GG19}. The Helmert sub-matrix is an orthonormal $(UC-1) \times UC$ matrix formed by deleting the first row of the Helmert orthogonal matrix (see \cite{G21} and \cite{TS22} for technical details). If we denote this Helmert sub-matrix as $\bm{H}$, then the ILR-transformed vector is defined as
\begin{equation}\label{eqn:ILR}
    \bm{z}(\bm{d}_t) = \bm{H} \bm{w}(\bm{d}_t),
\end{equation}
and is no longer subject to any constraint. That is $\bm{z}(\bm{d}_t) \in \mathcal{R}^{UC-1}$, and all of its elements can take any real value. 

The CLR and ILR aim to transform compositional data into real unconstrained space. On the other hand, as both these transformations are based on taking logarithms, then such methods will not work if one or more of the actual death counts, and subsequently one or more of the $d_{t,u,c}$'s, are exactly zero in value. This is the motivating problem for our subsequent developments as, in practice, many datasets of death counts tend to include zeros for some cause and age combinations. 

\subsection{The Lee-Carter model for compositional data}\label{sec:LC-CODA}

We describe a modification of the LC model introduced by \citet{O08} for compositional data. We refer to this model as the LC-CODA model, and its construction can be summarized in the following steps.
\renewcommand{\labelenumi}{(\Roman{enumi})}
\begin{enumerate}
\item Centre each row of $\bm{D}$ in~\eqref{eqn:D Matrix} by taking the inverse perturbation of the geometric mean from each row of death densities. This results in a matrix of centred death densities, denoted here as $\widetilde{\bm{D}}$.
\item Apply the CLR transformation to each row of $\widetilde{\bm{D}}$, mapping the vector of $UC$-compositions for a given year $t$ from the simplex to a $UC$-dimensional Euclidean subspace. 
\item Fit and forecast the transformed data using the LC model. Note other more sophisticated models are possible here \citep[e.g.,][]{BCO17, KEK+19, KEB+20}, and this step and all our developments can be modified to employ such approaches. For simplicity, though, we focus on the LC model.
\item Back-transform the estimated death densities to the simplex by inverting the CLR transformation and performing a compositional perturbation to the geometric mean for each row estimate to obtain the final forecasted compositional results.
\end{enumerate}

\sloppy We elaborate each of the steps above in detail. Consider the matrix of compositional death densities in~\eqref{eqn:D Matrix}, and compute $\mathbf{g}$ as the $UC$-vector, the elements of which are given by the column-wise geometric mean of $\mathbf{D}$, that is, 
\begin{math}
    \mathbf{g} = ( 
        (\prod_{t=1}^T d_{t, 1, 1})^{1/T}, (\prod_{t=1}^T d_{t, 1, 2})^{1/T}, \dots, (\prod_{t=1}^T d_{t, U, C})^{1/T} )
\end{math}.
Next, define the perturbation operation and its inverse as follows \citep{A82}. For two vectors of compositions $X = (x_1, x_2, \dots, x_n)$ and $Y = (y_1, y_2, \dots, y_n)$, all of the elements of which are non-zero,  we have
\begin{align*} 
    \text{Perturbation:} \quad &X \oplus Y = C \left( x_1y_1, x_2y_2, \dots, x_ny_n \right) \\
    \text{Inverse perturbation:} \quad &X \ominus Y = C \left( \frac{x_1}{y_1}, \frac{x_2}{y_2}, \dots, \frac{x_n}{y_n} \right),
\end{align*}
where the operator $C(\cdot)$ ``closes" the row, that is, normalizes by dividing each entry by the sum of all entries.

In Step (I) of fitting the LC-CODA model, we apply a centring process to construct a matrix of centred death densities, $\bm{\widetilde{D}}$, where the $t$\textsuperscript{th} row of $\bm{\widetilde{D}}$ for $t = 1,\ldots, T$ is computed as
\begin{equation}\label{eqn:centred vector}
    \widetilde{\bm{d}}_t = \bm{d}_t \ominus \mathbf{g} = C \left(
        \frac{d_{t, 1, 1}}{(\prod_{t=1}^T d_{t, 1, 1})^{1/T}}, \frac{d_{t, 1, 2}}{(\prod_{t=1}^T d_{t, 1, 2})^{1/T}}, \dots , \frac{d_{t, U, C}}{(\prod_{t=1}^T d_{t, U, C})^{1/T}} \right).
\end{equation}
Note the elements in $\mathbf{g}$ can be considered analogues of the age- and cause-specific average mortality over time in a standard LC model. 

In Step (II), we apply the CLR transformation to obtain the vector $\bm{w}(\widetilde{\bm{d}}_t) = (w(\widetilde{d}_{t,1,1}), w(\widetilde{d}_{t,1,2}),\ldots,w(\widetilde{d}_{t, U, C}))$ for $t=1,\ldots, T$, where to be clear the elements are computed using~\eqref{eqn:CLR} except replacing $d_{t,u,c}$ with $\widetilde{d}_{t,u,c} = d_{t,u,c}/(\prod_{t=1}^T d_{t,u,c})^{1/T}$.
Let $\bm{w}(\bm{\widetilde{D}})$ denote the resulting $T \times UC$ matrix formed by stacking the $\bm{w}(\widetilde{\bm{d}}_t)$'s as row vectors on top of one another.

In Step (III), we apply the singular value decomposition to $\bm{w}(\widetilde{\bm{D}})$ and estimate the Lee-Carter mortality model analogous to how it is done for the non-compositional setting. We provide details of this in Appendix \ref{sec:LC}, but to summarise, we fit a model of the form 
\begin{equation}\label{eqn:LC CODA CLR}
w(\widetilde{d}_{t, u, c}) = b_{u, c}k_{t, c} + \epsilon_{t, u, c},
\end{equation}
where $b_{u, c}$ denotes age- and cause-specific coefficients that vary over time, $k_{t, c}$ denotes factors of time-varying indices for the level of mortality, and $\epsilon_{t, u, c}$ denotes a residual error term. Note that a mean/intercept term is omitted from~\eqref{eqn:LC CODA CLR} due to centring from the geometric mean in Step (I). For forecasting, we can adopt a similar approach to \citet{KEK+19} and \citet{ZHH23}, among others, who applied time series methods such as random walk with drift to $k_{t, c}$, and substitute forecasted values of these back in to~\eqref{eqn:LC CODA CLR}.

Finally, in Step (IV), after obtaining forecasted values of $w(\widetilde{d}_{t, u, c})$, we can apply an inverse CLR transformation followed by a perturbation operation to obtain the actual forecasted death density distribution. In detail, suppose that at future time $T' > T$, the predicted value of the time factor for cause $c$ is given by $\widehat{k}_{T',c}$, while the estimated age- and cause-specific coefficients from Step (III) are given by $\widehat{b}_{u,c}$. Then a vector of forecasted centred death densities is given by $\widetilde{\bm{d}}_{T'} = (\widetilde{d}_{T',1,1},\widetilde{d}_{T',1,2},\ldots,\widetilde{d}_{T',U,C})$ where $\widetilde{d}_{T',u,c} = w^{-1}(\widehat{b}_{u,c}\widehat{k}_{T',c})$ and $w^{-1}(\cdot)$ denotes the inverse CLR transformation. The corresponding vector for forecasted death densities from the LC-CODA model is then given by $\widehat{\bm{d}}_{T'} = \widetilde{\bm{d}}_{T'} \oplus \bm{g}$.

Compared to modelling mortality rates independently, one key element of using the LC-CODA model is that death counts are naturally redistributed through compositional constraints. As mortality changes over time, if some deaths do not occur at a specific age band and cause, they are naturally shifted towards a different age band and cause group. This maintains subcompostional coherence with the total number of deaths per year as given by the initial life table and ensures the disaggregated death forecasts will be coherent with the overall aggregated mortality forecast \citep{O08}. In the context of compositional data, subcompositional coherence refers to the property that relationships between parts of a composition are unaffected by forming subcompositions, such that results and summary statistics based on the subcomposition are the same as the composition \citep{G21}. On the other hand, due to its reliance on the CLR transformation, the LC-CODA model is unable to handle zero values in the raw densities $d_{t,u,c}$, and these would need to be omitted, aggregated, or replaced with an arbitrarily small value before step (I).

We present a more detailed exposition of LC-CODA in Appendix~\ref{sec:LC}, which we use in the application of the $\alpha$-transformation and log-ratio transformations in this paper.

\section{Mortality by Cause Using the \texorpdfstring{$\alpha$}{alpha}-Transformation }\label{sec:alpha transformation}

Motivated by the challenges of applying LRA to cause-of-death mortality modelling where there are one or more zero values in the death densities, we propose using the $\alpha$-transformation before applying the LC-CODA model for forecasting. 

The $\alpha$-transformation can be viewed as a Box-Cox transformation applied to the ratios of components, where $\alpha \in (0,1]$ is a tuning parameter that is tuned to handle compositional challenges in the data with zeros \citep{TPW11}. In detail, let $w^\alpha(x)$ represent the Box-Cox transform of a random variable $x$ \citep{BC64},
\begin{equation*}
    w^\alpha(x) = \begin{cases} \ln(x) &  \alpha = 0 \\
    \frac{x^\alpha -1}{\alpha} &  \alpha \neq 0,
    \end{cases}
\end{equation*}
and recall the matrix of centred death densities $\bm{\widetilde{D}}$ in~\eqref{eqn:centred vector}. For row $t = 1,\ldots,T$, the $\alpha$-transformation is then defined as:
\begin{equation}\label{eqn:alpha}
    \bm{z}^\alpha(\bm{\widetilde{d}}_t) = \bm{H} \bm{w}^\alpha(\bm{\widetilde{d}}_t),
\end{equation}
where $\bm{H}$ is the Helmert sub-matrix defined as part of the ILR transformation in~\eqref{eqn:ILR}, and $\bm{w}^\alpha (\bm{\widetilde{d}}_t)$ denotes the vector where the Box-Cox transformation is applied to each element of $\widetilde{\bm{d}}_t$. That is, $w^\alpha(\widetilde{d}_{t,u,c}) = \ln(\widetilde{d}_{t,u,c})$ if $\alpha = 0$, otherwise $w^\alpha(\widetilde{d}_{t,u,c}) = (UC)(\widetilde{d}_{t,u,c}^\alpha -1)/\alpha$ for $\alpha \neq 0$. Note when $\alpha = 0$, the transformation reduces to the ILR transformation defined in~\eqref{eqn:ILR}. If there is no left matrix multiplication by the Helmert sub-matrix $H$, then we obtain the CLR in~\eqref{eqn:CLR}. Critically, when $\alpha$ is restricted to be greater than zero, the transformed values are well defined even when the raw death densities $\widetilde{d}_{t,u,c} = 0$. This differs from both the ILR and CLR, neither of which can be computed for zero values.

The corresponding sample space of the $\alpha$-transformation is known as the $\alpha$ space, which we denote as $\mathbb{A}^{UC-1}_\alpha$ and is given by
\begin{equation*}
    \mathbb{A}^{UC-1}_\alpha = \left\{ \bm{z}^\alpha(\bm{\widetilde{d}}_t) \middle| -\frac{1}{\alpha} \leq w^\alpha(\widetilde{d}_{t,u,c}) \leq \frac{(UC-1)}{\alpha}, \quad \sum_{u=1}^{U}\sum_{c=1}^{C} w^\alpha(\widetilde{d}_{t,u,c}) = 0 \right\}.
\end{equation*}
It is not difficult to see that, similar to the ILR transformation, the vectors in $\mathbb{A}^{UC-1}_\alpha$ are not subject to the zero-sum constraint. As $\alpha \rightarrow 0$, then $\mathbb{A}^{UC-1}_\alpha$ tends to the $(UC-1)$ dimensional real space $\mathcal{R}^{UC-1}$; this is again consistent with the ILR, except now zero values of death densities can be handled provided $\alpha \ne 0$ \citep{TS22}. On the other hand, when $\alpha = 1$, the $\alpha$-transformation is equivalent to RDA, that is, the same as applying standard multivariate analysis ignoring the compositional constraint. While $\alpha$ is often determined using a data-driven approach through maximum likelihood estimation \citep{TPW11}, for strong forecasting performance, in Section~\ref{sec:alpha-optimisation}, we discuss an alternative method based on minimising out-of-sample prediction accuracy.

To construct the Lee-Carter model in conjunction with the $\alpha$-transformation, we can apply similar steps to those discussed in Section~\ref{sec:LC-CODA}, except that Step (II) is modified to Step (IIa) where we apply the $\alpha$-transformation instead of the CLR, and Step (IV) is modified to Step (IVa) where the transformation back to the simplex requires inverting the $\alpha$-transformation to obtain the final forecast. With regards to the latter, after forecasting the factors $k_{t,c}$ in a similar manner to Section \ref{sec:LC-CODA}, the forecast result derived based on the $\alpha$-transformed data needs to be mapped back to the compositional simplex.

In detail, at future time $T^{'}>T$ and for $\alpha > 0$, let $\widehat{\bm{z}}^\alpha(\widetilde{\bm{d}}_{T^{'}}) = (\widehat{z}^{\alpha}(\widetilde{d}_{T',1,1})$,$\ldots,\widehat{z}^{\alpha}(\widetilde{d}_{T',U,C}))$ denote the vector of forecasted $\alpha$-transformed centred death densities, where $\widehat{z}^\alpha(\widetilde{d}_{T',u,c}) = \widehat{b}_{u,c}\widehat{k}_{T',c}$. Then, the vector of corresponding inverse $\alpha$-transformed values is given by
$    v^\alpha(\bm{\widetilde{d}}_{T'}) 
    = \alpha \bm{H}^\top \widehat{\bm{z}}^\alpha(\widetilde{\bm{d}}_{T'}) + 1$.

Afterwards, the forecast vector of death densities at time $T'$ is given by
\begin{equation*} 
    \bm{\widetilde{d}}_{T'} = \left( \frac{v^{1/\alpha} (\bm{\widetilde{d}}_{t, 1, 1})}{\sum_{u=1}^{U} \sum_{c=1}^{C} {v^{1/\alpha} (\bm{\widetilde{d}}_{T', u, c})}} , \cdots , \frac{v^{1/\alpha} (\bm{\widetilde{d}}_{t, U, C})}{\sum_{j=1}^{U}\sum_{k=1}^{C}{v^{1/\alpha} (\bm{\widetilde{d}}_{T, u, c})}} \right),
\end{equation*}
and $\widehat{\bm{d}}_{T'} = \widetilde{\bm{d}}_{T'} \oplus \bm{g}$.

To conclude, we remark that as long as the forecasted data $\widehat{\bm{z}}^\alpha(\widetilde{\bm{d}}_{T'})$ lies inside $\mathbb{A}_\alpha^{UC-1}$ defined by the original data, then it can be mapped back to the simplex for inference. In some cases during the process of forecasting, for example, for long-term forecasts when $T' \gg T$, it is possible one or more values of $\widehat{\bm{z}}^\alpha(\widetilde{\bm{d}}_{T'})$ are less than $-1/\alpha$ and lie outside the $\alpha$-space. This indicates the corresponding forecasts are at or crossing the boundary of the simplex. In such cases, to ensure the inverse $\alpha$-transformation is possible, we choose to set corresponding elements of $\widehat{\bm{z}}^\alpha(\widetilde{\bm{d}}_{T'})$ equal to the boundary value of $-1/\alpha$ \citep[see, e.g.][for a similar treatment]{TPW11}.

\section{Application to the Human Cause-of-Death Database} \label{sec:application}

We illustrate an application of the $\alpha$-transformation coupled with an LC model to cause-of-death counts and life-table deaths for two data sets from England and Wales and the US as part of the \citet[HCD,][]{HMD24}. England and Wales were selected as there have been relatively minimal fluctuations in cause composition during the available data period, while the US was selected to assess the performance of the proposed $\alpha$-transformation for a larger data set spanning more historical years. Disaggregated causes of death within the cardiovascular causes were selected since cardiovascular disease has been steadily decreasing over the past few decades but remains the second-largest cause of death in the UK \citep{BHF23, CKS23, RJW22}. Data on the complete list of causes of death were obtained, containing 103 causes at the ``long" level for England and Wales, and 206 causes for US death counts. We treat males and females as separate data sources, and perform analysis separately by gender; this is consistent with treatment in earlier CODA literature \citep[e.g.,][]{O08, KEK+19}.

To perform analysis and forecasting, we aggregated based on age bands and selected causes. We constructed nine age bands: ages~0 to~24, ages~25 to~34, ages~35 to~44, ages~45 to~54, ages~55 to~64, ages~65 to~74, ages~75 to~84, ages~85 to~95, and ages over~95. There was an additional age band for the US data comparison, namely ages~90 to~99 and then ages over~100. This additional age band was possible due to the availability of the granular death count data from \cite{HMD24} for the US. Note the age band 0 to 24 is not a homogeneous group relative to the other age bands, but the reason for aggregating at these ages is twofold: first, for application to life insurance, analysis is typically performed for working age groups; second, by aggregating across 0 to 24  there is greater credibility in death counts. We leave the assessment of the variation of deaths by cause at younger ages as an avenue for future investigation. 

Turning to causes, for England and Wales death counts, we aggregated death counts by cause into 11 causes as per the HCD shortlist, with only the cardiovascular causes disaggregated to the ``long" list level. For the US death counts, we ensured the same ICD-10 causes of death were used for comparison. These same cardiovascular causes were mapped to 12 causes as per the HCD ``long" list level for the US data. Cardiovascular causes were selected as cardiovascular disease causes of death have steadily decreased over the data period, as introduced at the start of this section. All other causes of death were grouped and aggregated for analysis. The selected cardiovascular causes of death for both datasets are shown in Table~\ref{tab:usuk_causes}. 

\begin{table}[!htb]
\centering
\tabcolsep 0.2in
\begin{tabular}{@{}m{9.5cm}m{5cm}@{}}
\toprule[1.5pt]
         \textbf{Mortality causes at the ``long" level} & \textbf{ICD-10 causes of death} \\
         \midrule
\multicolumn{2}{c}{\underline{England and Wales data}} \\
         48: Rheumatic heart disease & I00--I09 \\
         49: Essential hypertension & I10 \\
         50: Hypertensive disease (heart, kidney, secondary) & I11--I15 \\
         51: Acute myocardial infarction & I21--I23 \\
         52: Other IHD & I20, I24, I25 \\
         53: Pulmonary heart diseases & I26--I28 \\
         54: Non-rheumatic valve disorders & I34--I38 \\
         55: Cardiac arrest & I46 \\
         56: Heart failure & I50 \\
         57: Other heart diseases & I30--I33, I40--I45, I47--I49, I51 \\
         1: All other causes of death & All other ICD--10 \\\\
\multicolumn{2}{c}{\underline{US data}} \\
         102: Acute Rheumatic & I00--I02 \\
         103: Chronic Rheumatic & I05--I09 \\
         104: Hypertension & I10 \\
         105: Hypertensive (heart) & I11 \\
         106: Hypertensive (renal) & I12 \\
         107: Hypertensive (both heart and renal) & I13 \\
         108: Myocardial Infarction & I21 \\
         109: IHD acute & I20, I24 \\
         110: IHD chronic & I25 \\
         111: Pulmonary & I26--I28 \\
         112: Other cardiovascular causes of death & I30--I51 \\
         1: All other causes of death & All other ICD--10 \\
\bottomrule[1.5pt]
\end{tabular}
\caption{Selected causes of death, disaggregated for cardiovascular causes, used in our application to England and Wales data (top) and US data (bottom) from \cite{HMD24}.}\label{tab:usuk_causes}
\end{table}

In the disaggregated data for cardiovascular deaths, zero death counts were present across most causes in the disaggregated cardiovascular death category over the available period (2001 to 2016 for England and Wales, and 1979 to 2021 for US,) and when split by age band and across both genders. For example, for England and Wales male data, rheumatic heart disease had zero counts for ages less than 20 (and also for ages 20 to 30 in 2010) for 2002, 2006, 2010, and 2012 to 2015. Also, for males, cardiac arrest death counts were zero for ages 40 to 50 in the year 2004. Similarly, for US male data, acute rheumatic deaths had zero counts for ages less than 20 in 1998, 2002, 2004, 2006 -- 2009, 2014 -- 2016, 2018 -- 2019, and 2021. The same cause had zero counts for ages up to 40 in 2007, and across other older bands in the available years.

\begin{figure}[!htb]
\centering
\includegraphics[width=7.9cm]{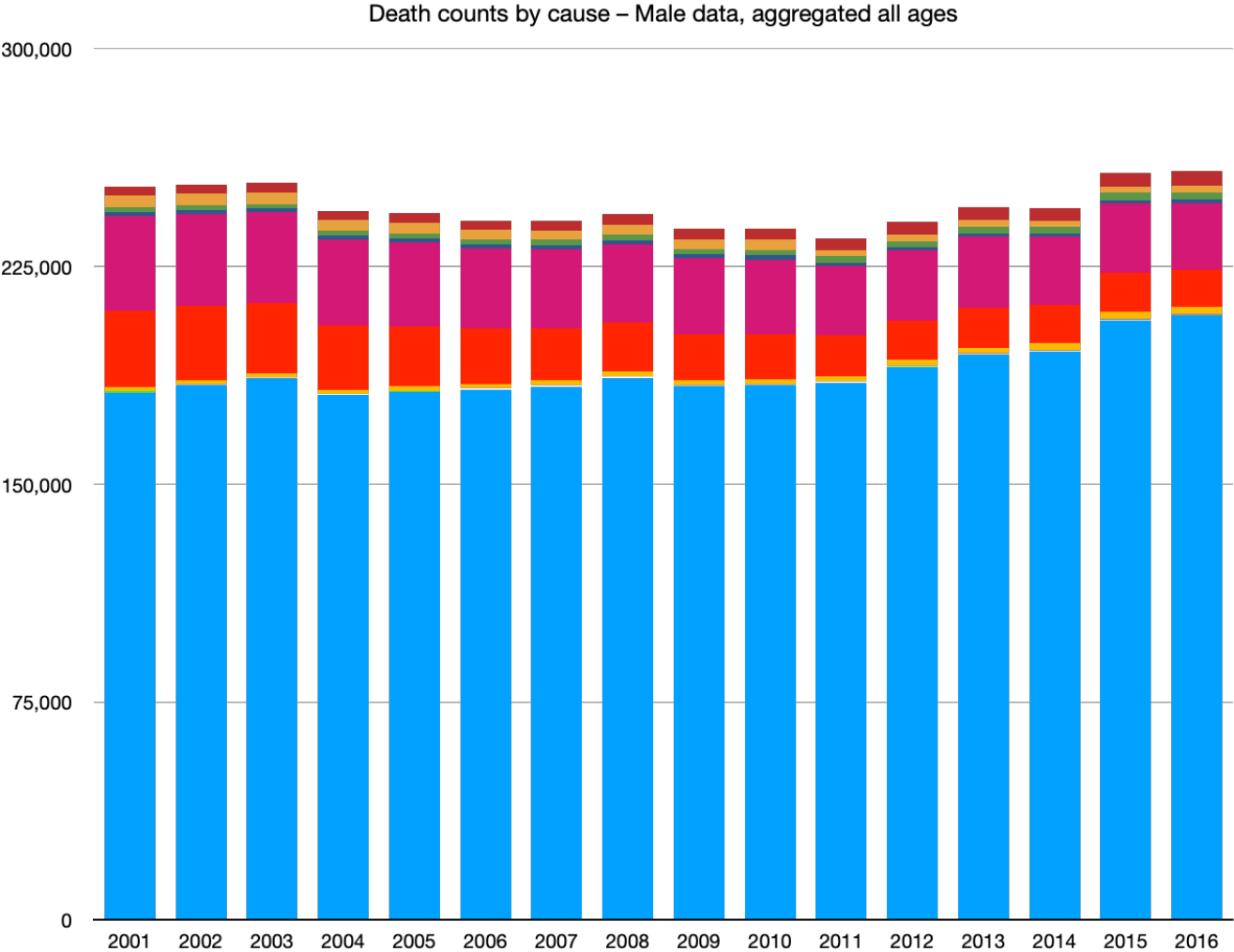}
\includegraphics[width=7.9cm]{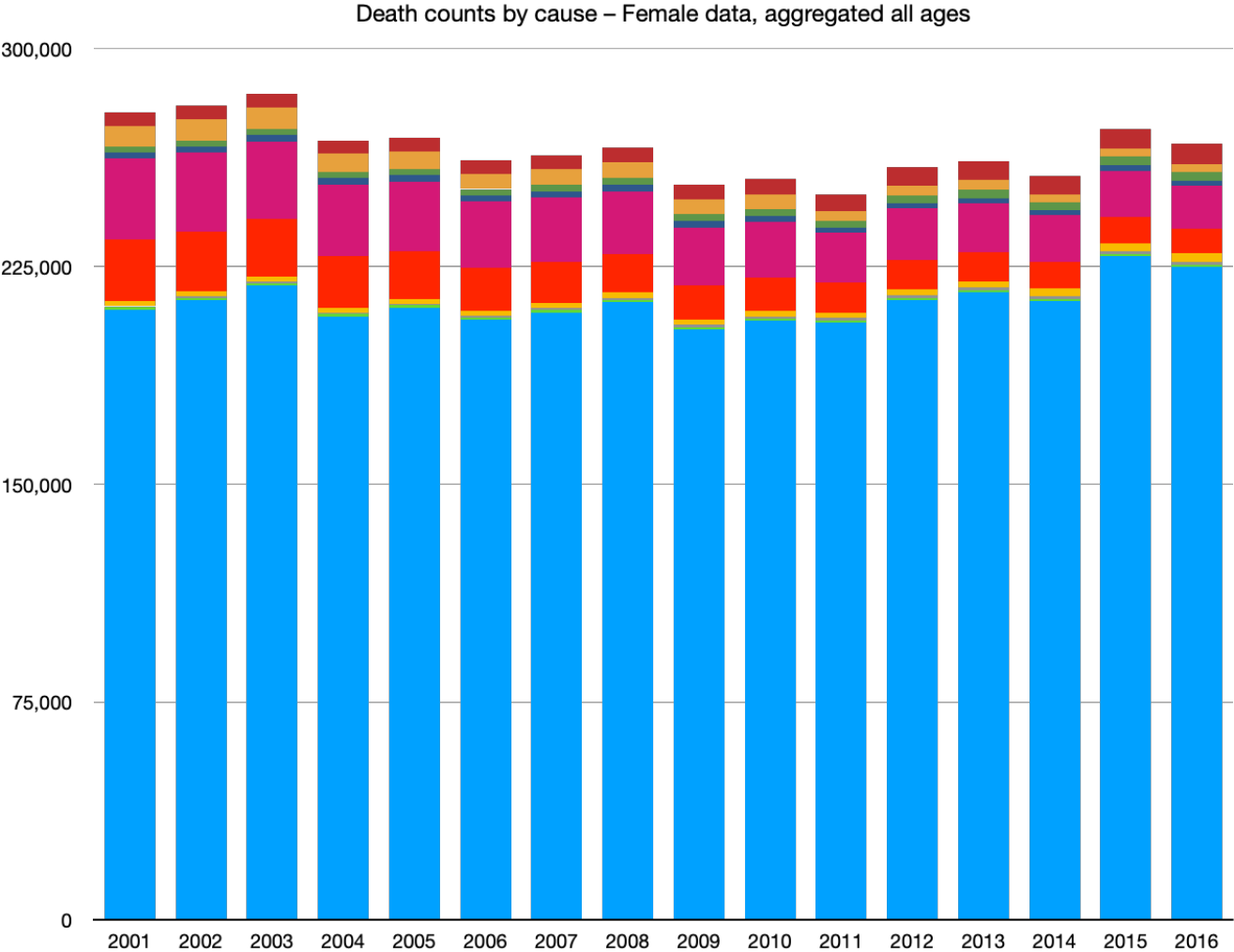}
\includegraphics[width=7.9cm]{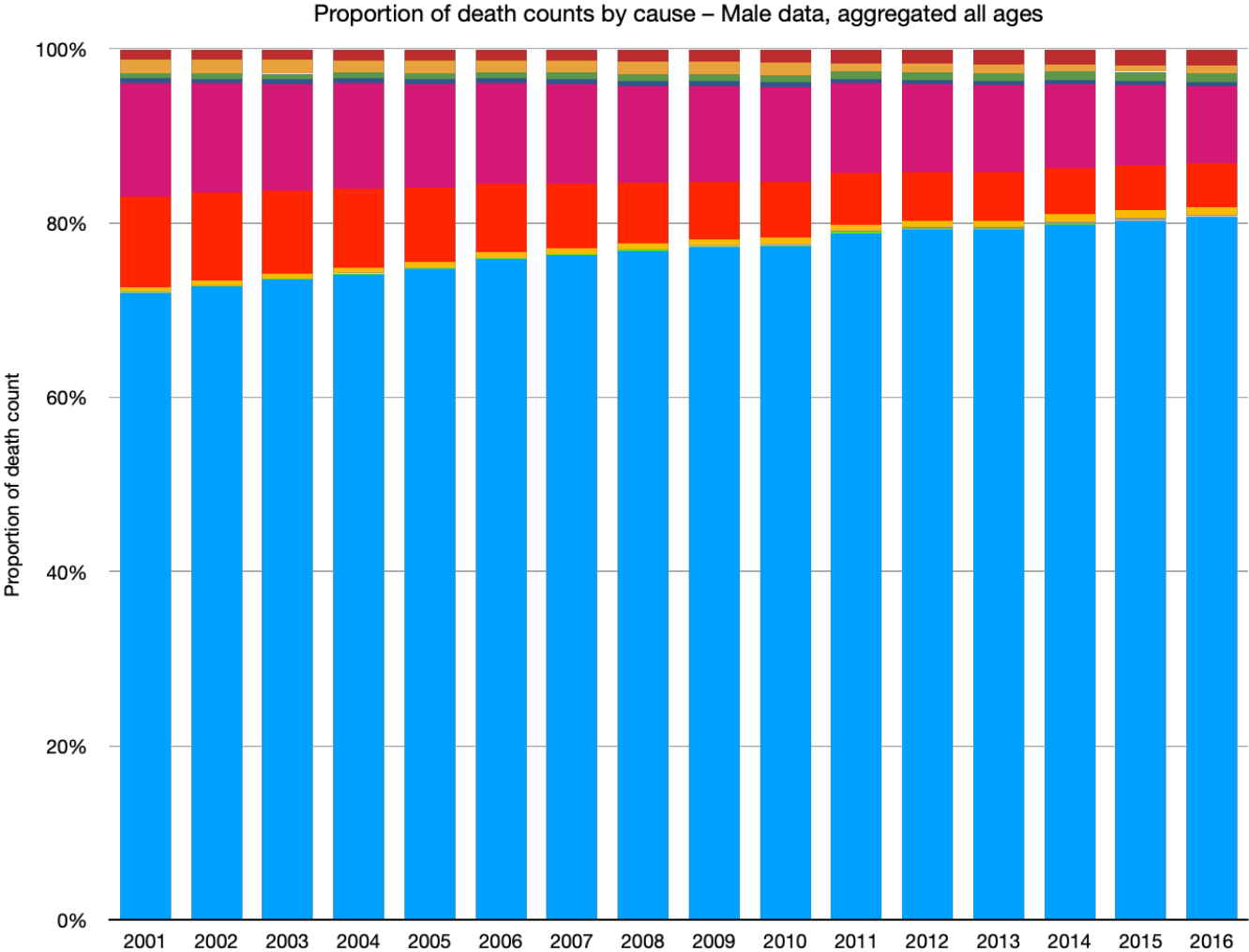}
\includegraphics[width=7.9cm]{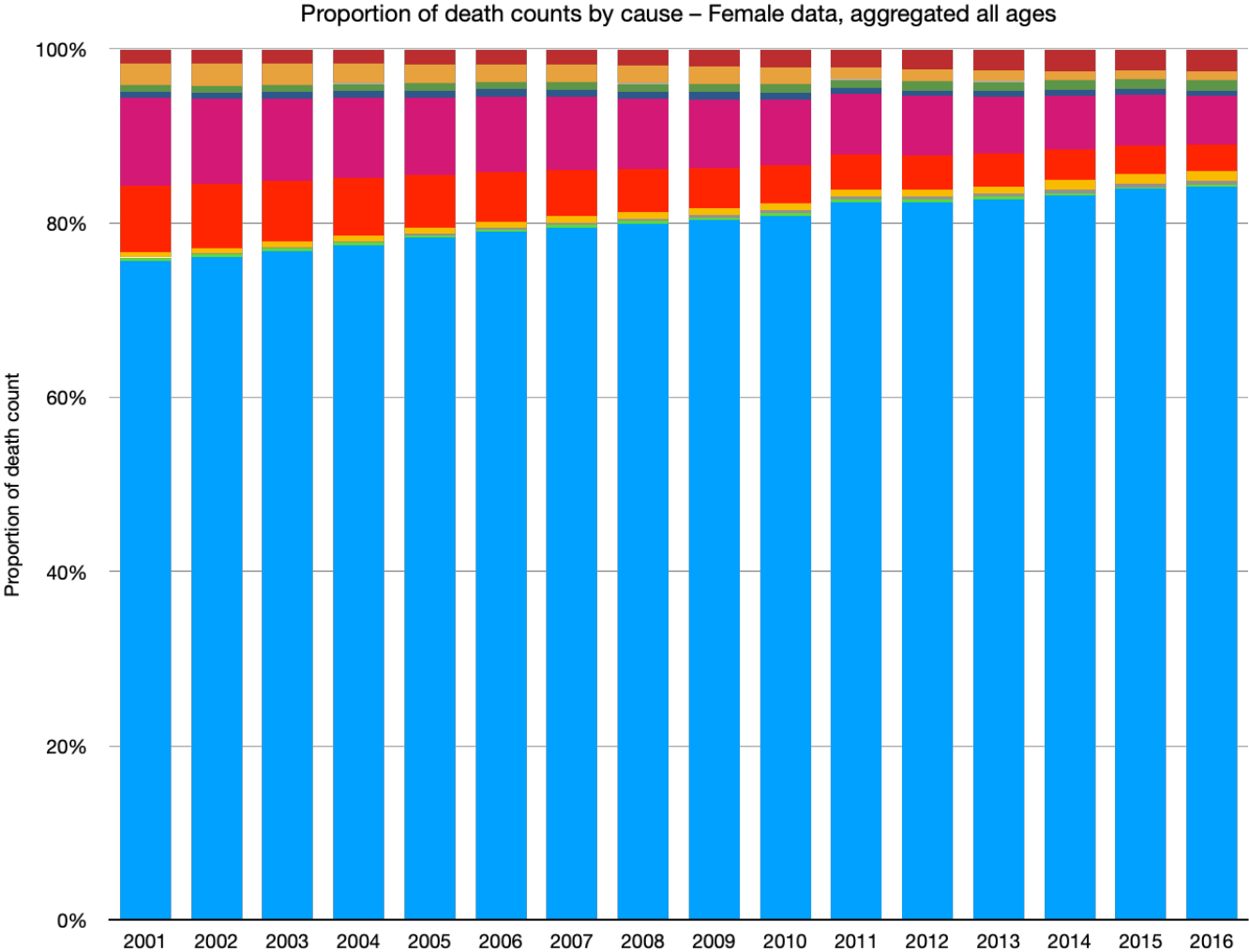}
\quad
{\includegraphics[width=15cm]{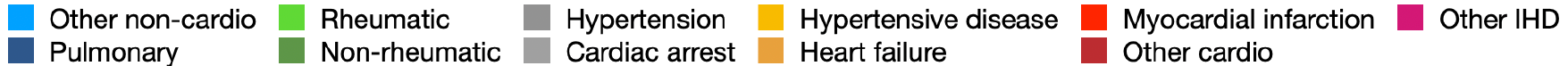}}
\caption{Death counts by cause for England and Wales deaths from 2001 to 2016. The top row presents death counts by cause (disaggregated cardiovascular causes) for males (left) and females (right) in our application to England and Wales data from \cite{HMD24}. The bottom row presents the same data but converted to the composition of cardiovascular deaths by cause.} \label{fig:cardio_countsandprop}
\end{figure}

\begin{figure}[!htb]
\centering
\includegraphics[width=7.9cm]{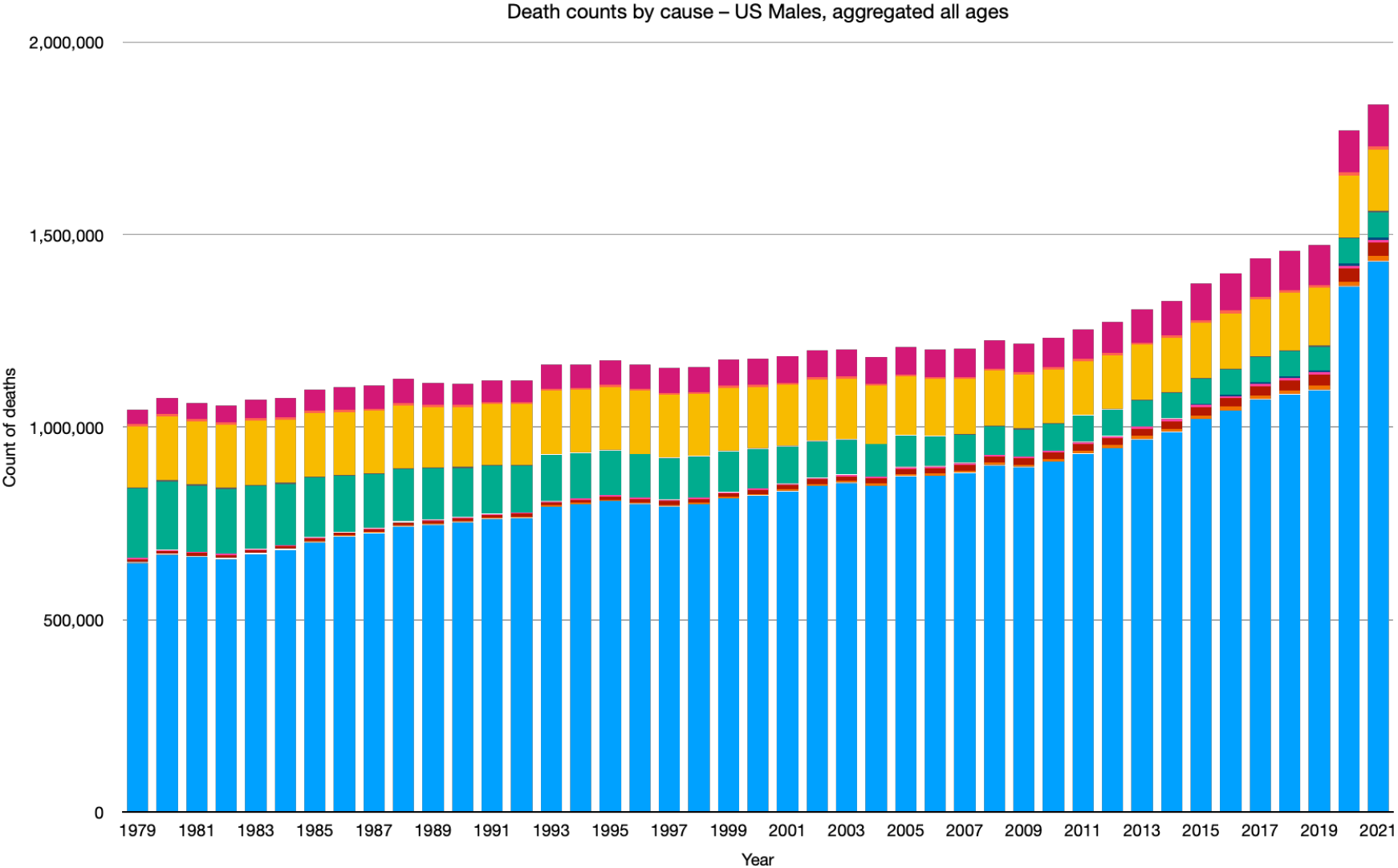}
\includegraphics[width=7.9cm]{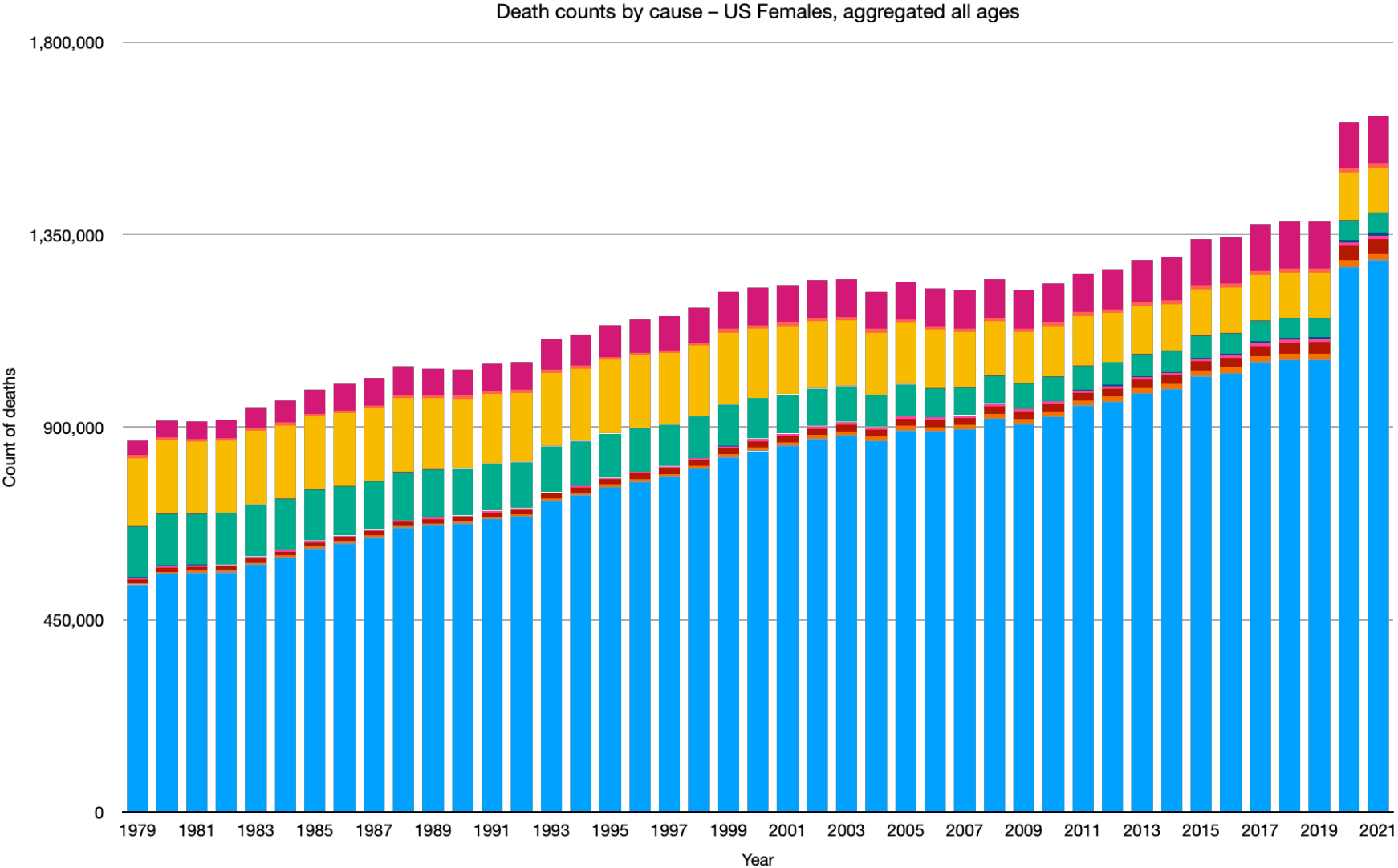}
\includegraphics[width=7.9cm]{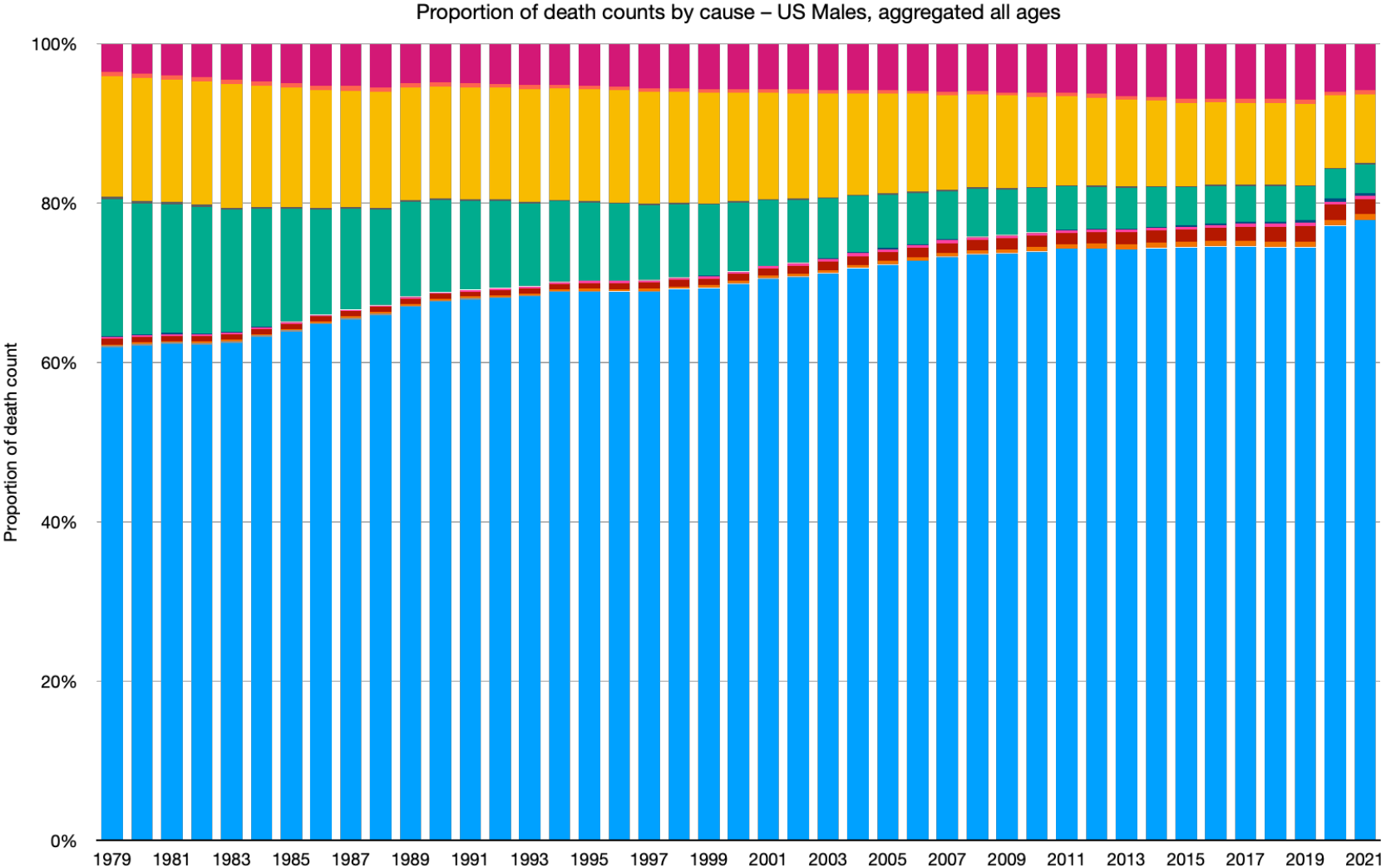}
\includegraphics[width=7.9cm]{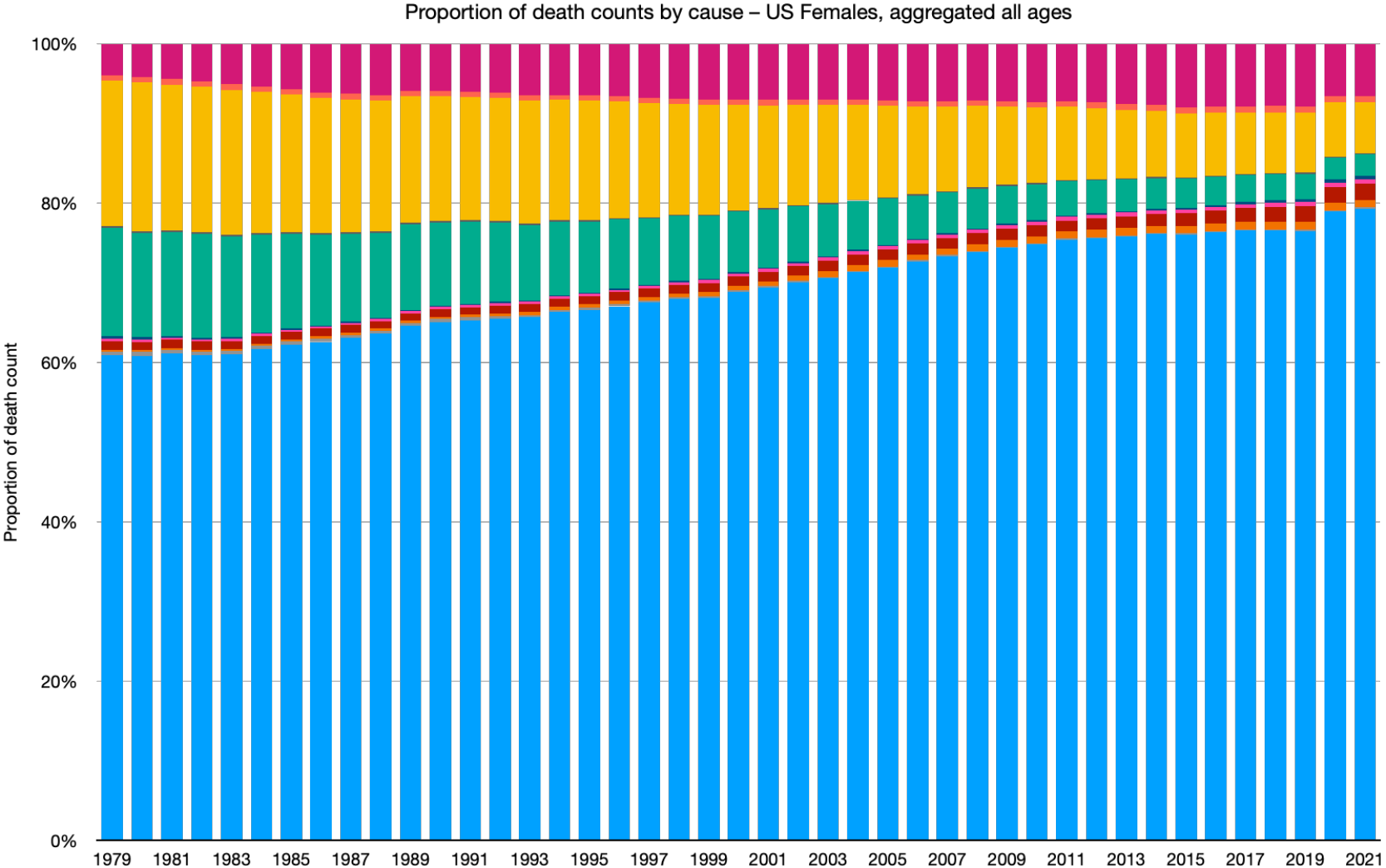}
\quad
{\includegraphics[width=15cm]{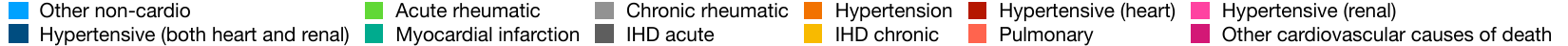}}
\caption{Death counts by cause for US deaths from 1979 to 2021. Note the two years 2020 and 2021 show a spike in deaths, likely due to COVID. The top row presents death counts by cause (disaggregated cardiovascular causes) for males (left) and females (right) in our application to England and Wales data from \cite{HMD24}. The bottom row presents the same data but converted to the composition of cardiovascular deaths by cause.} \label{fig:us_cardio_countsandprop}
\end{figure}

In total, for the England and Wales death counts, of the ten cardiovascular causes of death, six had one or more zero counts across both genders in the data: rheumatic heart disease, essential hypertension, hypertensive disease, acute myocardial infarction, cardiac arrest, and heart failure. Not surprisingly, zero death counts for most causes tended to be more prevalent in some years at younger ages (below 50). Similarly, for US death counts, five of the total 11 cardiovascular causes had zero counts across genders and age bands: acute and chronic rheumatic, hypertension, and hypertensive (both heart and renal). Figures~\ref{fig:cardio_countsandprop} {and~\ref{fig:us_cardio_countsandprop}} presents aggregated death counts across all ages from 2001 to 2016 for England and Wales, and from 1979 to 2021 for US deaths. As observed, the number of deaths for some causes is small, even when aggregated across all ages. With the above in mind, we anticipate forecast performance will improve by explicitly working with actual death counts, that is, including zero values, compared with the standard approach of excluding zeros or replacing them with an arbitrarily small amount.

\subsection{Tuning \texorpdfstring{$\alpha$}{alpha} parameter}\label{sec:alpha-optimisation}

To predict cause-of-death data with zero death counts, we proposed selecting an optimal value of $\alpha$ based on out-of-sample forecast accuracy as assessed via an expanding window cross-validation approach. Specifically, for the England and Wales data, as the available data only spanned 16 years, we adopted a simple four-fold expanding window. For the US data, as there was 43 years of data, we adopted a ten-fold expanding window. On England and Wales deaths, this meant the first fold consists of years 2001--2008 for training and 2009--2012 for validation, the second fold consisted of 2001-2009 for training and 2010--2012 for validation (i.e., the training window was increased by one year), and so on. In each fold, the $\alpha$-transformation coupled with the LC model as detailed in Section \ref{sec:alpha transformation} was fitted to the training set and forecasts made to the validation set. The years 2013--2016 were held out from all four folds as a test set. Analogously, for the US data the first fold consisted of years 1979--2001 for training and 2002--2011 for validation, the second fold consisted of 1979-2002 for training and 2003--2011 for validation, and so on. The years 2012--2021 were held out from all folds as a test set. We remark that, as the compositional cause of death data exhibits a natural time-series dependence, then an expanding window (or forward chaining) cross validation method was adopted to tune $\alpha$; we refer to \citet{R00} and \citet{S19} for more details around cross validation in the context of time series analysis.

For both datasets, we selected $\alpha$ based on minimising either the average root mean square error (RMSE) or average mean absolute error (MAE) across the four validations sets,
\begin{align*}
    \text{RMSE}_k &= \sqrt{\frac{\sum_{t=1}^{T_k} \sum_{u=1}^9 \sum_{c=1}^{11} (\text{observed}_{t,u,c} - \text{predicted}_{t,u,c})^2}{N}}; \\  
    \text{MAE}_k &= \frac{\sum_{t=1}^{T_k} \sum_{u=1}^9 \sum_{c=1}^{11} \lvert \text{observed}_{t,u,c} - \text{predicted}_{t,u,c}\rvert}{N},
\end{align*}
\noindent where $\text{observed}_{t,u,c}$ generically denotes the death count for age band $u$, cause $c$ and the $t$\textsuperscript{th} year in the validation set, $\text{predicted}_{t,u,c}$ denotes the corresponding predicted death count, and $T_k$ denotes the number of years in the $k$\textsuperscript{th} validation fold. 

Both RMSE and MAE are widely used in model evaluation to measure forecast accuracy \citep{CD14, H22}. 
\begin{align*}
    \text{RMSE} = \frac{1}{4} \times \sum_{k=1}^{4} \text{RMSE}_k \\ 
    \text{MAE} = \frac{1}{4} \times \sum_{k=1}^{4} \text{MAE}_k
\end{align*}

Full results from applying the above cross-validation approach are provided in Appendix~\ref{subsec:tuningalpha_additionalresults}. Overall, the optimal $\alpha$ determined using the above cross-validation approach was $0.1$ and $0.8$ for males and females, respectively, when applied to England and Wales cause-of-death data. On the other hand, optimising $\alpha$ on the US data yielded values of $0.7$ and $0.9$, respectively, for males and females. In three of the four cases for optimising $\alpha$, the minimum RMSE and MAE produced the same results. Interestingly, the optimal $\alpha$ chosen for the US female data was $1.0$ when using RMSE as the criteria: since the $\alpha$-transformation here converges to RDA, this suggests the compositional constraint impacted the analysis to a lesser extent for this setting. On the other hand, since using MAE produced both lower RMSE and MAE in the validation sets compared with the optimal $\alpha$ determined using RMSE, then we decided to choose the optimal $\alpha$ as $0.9$ for the US female data.

\subsection{Results: England and Wales data}\label{sec:cardio_results} 

Using the values of $\alpha$ tuned in Section~\ref{sec:alpha-optimisation}, we produced mortality forecasts of proportions of deaths by cause for the test set (England and Wales years 2016--2020 and US years 2012--2021) using the $\alpha$-transformation coupled with the LC model. We compared this with several LRA methods in the literature for addressing zeros counts, including the CLR and ILR transformations where zeros were omitted from the data and the CLR and ILR with all zeros replaced by 0.25 or 0.5 before modelling. These additional methods were coupled with an LC model for forecasting and comparison. 

\begin{figure}[!htb]
\centering
{\includegraphics[width=15.5cm]{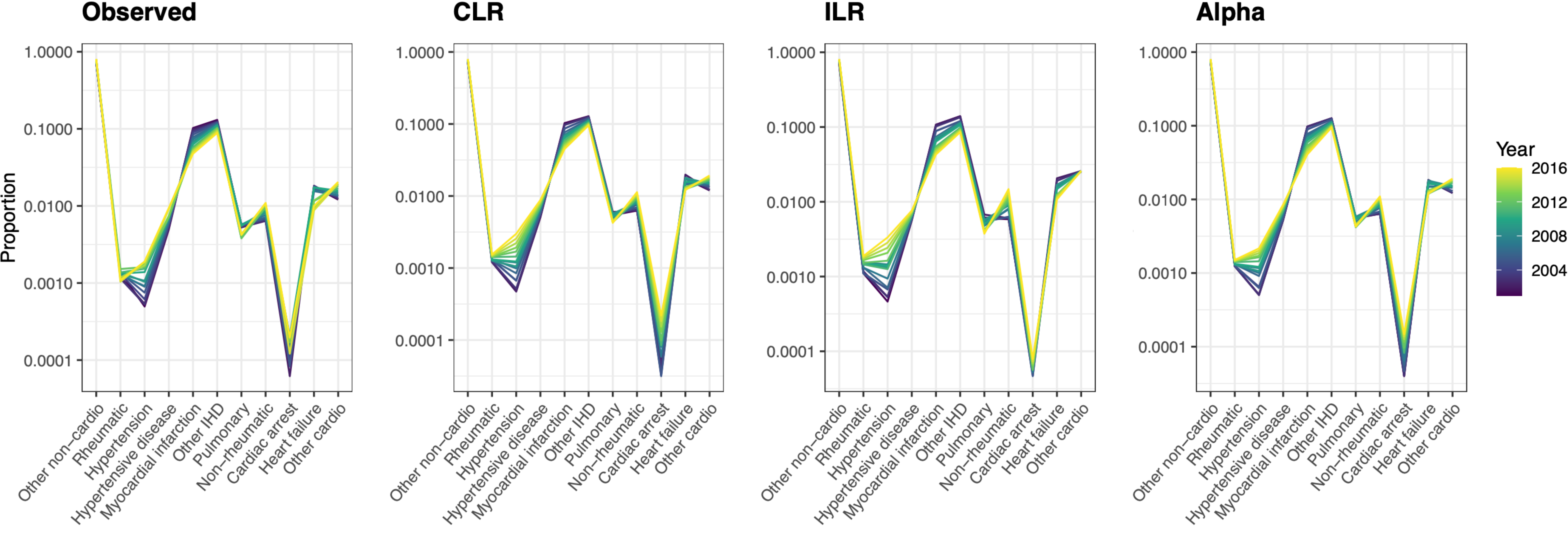}\label{fig:cardio_compare_m}}
\quad
{\includegraphics[width=15.5cm]{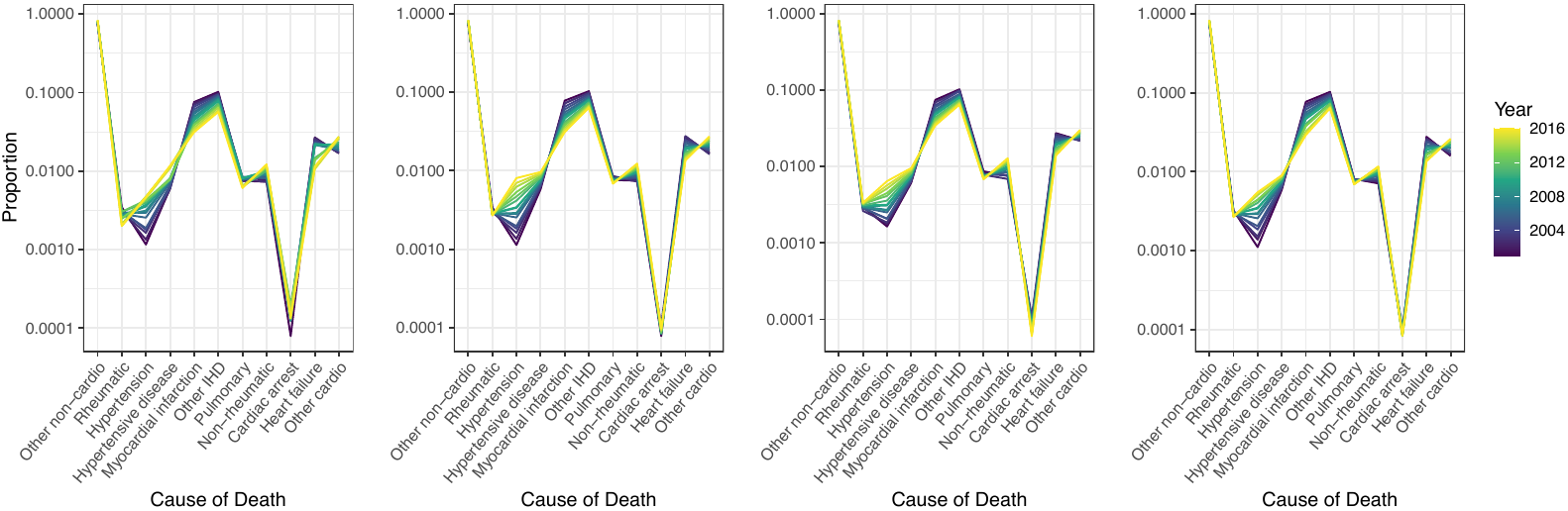}\label{fig:cardio_compare_f}}
\quad
\caption{Male (top row) and female (bottom row) mortality by cause in our application to England and Wales data from \cite{HMD24}, disaggregated for cardiovascular causes of death. The figures show the movement in actual proportion of deaths for each cause from 2001 to 2016 (left column), while the remaining three columns present results from applying CLR, ILR (with zeros removed), and $\alpha$-transformations, respectively.} \label{fig:cardio_compare}
\end{figure}

Table~\ref{tab:cardio_results} summarises the performance for females and males. Aside from the optimal values of $\alpha$, we also considered values $\alpha = 0.5$, $\alpha = 0.7$, $\alpha = 0.9$, and $\alpha = 1$, the latter equivalent to RDA, that is, ignoring the compositional constraint. The $\alpha$-transformation, on the whole, tended to produce better forecasting accuracy for the US data set compared with the CLR and ILR plus either \emph{ad-hoc} method of handling zero values. Improvements in the forecast were more evident when assessing MAE across both genders, although even with RMSE, the $\alpha$-transformation was the second-best performer. Visually, Figure~\ref{fig:cardio_compare} corroborates the results for males and females, where the $\alpha$-transformation better fits the observed data when compared with the corresponding CLR and ILR transformations.

\begin{table}[!htb] 
  \caption{\small{Forecast performance on test data, applying CLR, ILR, and the $\alpha$-transformation coupled with the LC model to England and Wales data from \cite{HMD24}, disaggregated for cardiovascular causes of death. For each metric and gender, the bolded values correspond to the error using optimal values of $\alpha$ tuned based on cross-validation. In contrast, underlined values correspond to the lowest metric in the out-of-sample forecast.}}
  \label{tab:cardio_results} 
\centering
\tabcolsep 0.27in
\begin{tabular}{@{}lllll@{}} 
 \toprule[1.5pt]
Method & \multicolumn{2}{c}{RMSE $\times$ 100} & \multicolumn{2}{c}{MAE $\times$ 100}  \\ [0.5ex] 
        & Male & Female & Male & Female \\
\cmidrule{1-5}
CLR (zeros omitted) & \underline{0.1777} & 0.2125 & 0.1030 & 0.1172\\
CLR (0.25 zero replacement) & 0.2311 & 0.3225 & 0.1154 & 0.1740\\
CLR (0.5 zero replacement) & 0.1892 & 0.2603 & 0.0980 & 0.1373\\
ILR (zeros omitted) & \underline{0.1777} & 0.2125 & 0.1030 & 0.1172\\
ILR (0.25 zero replacement) & 0.2311 & 0.3225 & 0.1154 & 0.1740\\
ILR (0.5 zero replacement) & 0.1892 & 0.2603 & 0.0980 & 0.1373\\
$\alpha$ = 0.1 & \bf{0.1818} & 0.2023 & \bf{0.1046} & 0.1121\\ 
$\alpha$ = 0.5 & 0.1852 & 0.1714 & \underline{0.0959} & 0.1011\\ 
$\alpha$ = 0.7 & 0.2109 & 0.1642 & 0.1064 & \underline{0.0994} \\
$\alpha$ = 0.8 & 0.2296 & \underline{\bf{0.1631}} & 0.1138 & \bf{0.0998}\\
$\alpha$ = 0.9 & 0.2526 & 0.1640 & 0.1228 & 0.1004\\
$\alpha$ = 1 (RDA) & 0.2809 & 0.1669 & 0.1329 & 0.1015\\ 
\bottomrule[1.5pt]
\end{tabular}
\end{table}

\begin{figure}[!htb]

{\includegraphics[width=0.82\textwidth]{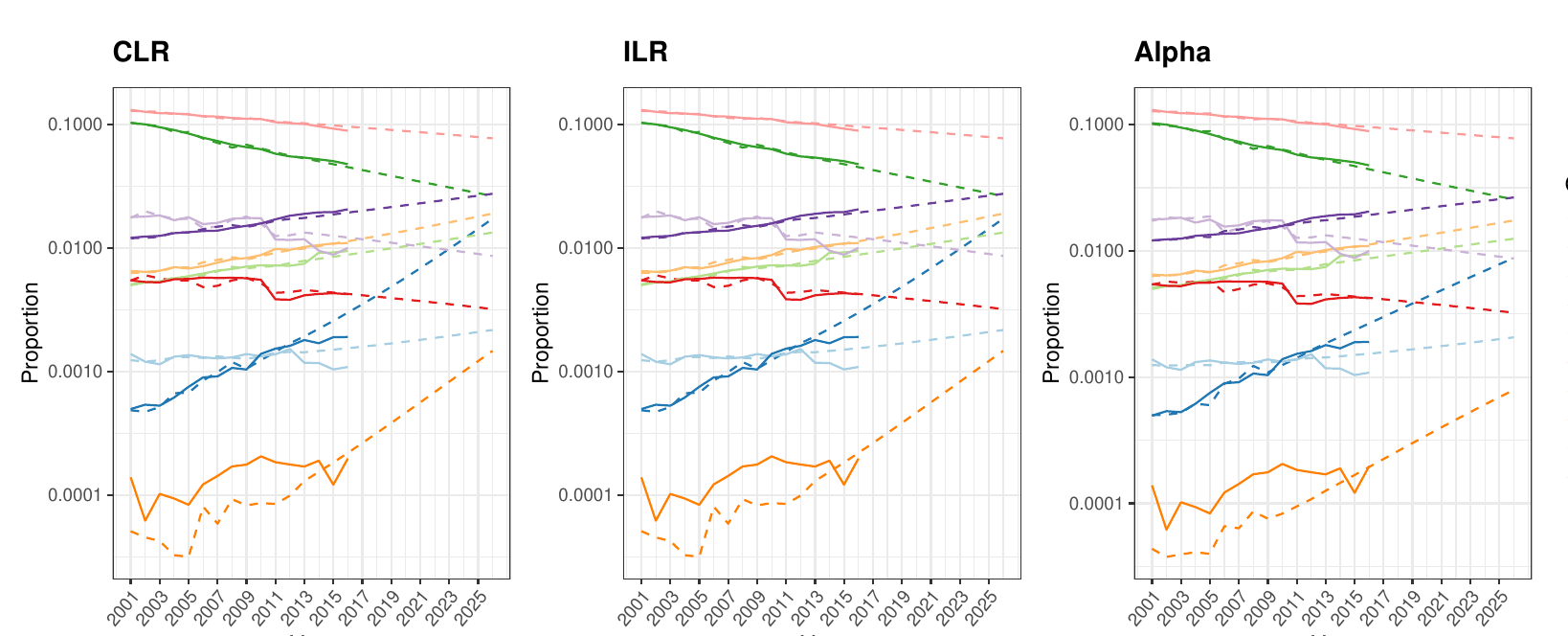}\label{fig:all_proj_m}}
\\
{\includegraphics[width=0.95\textwidth]{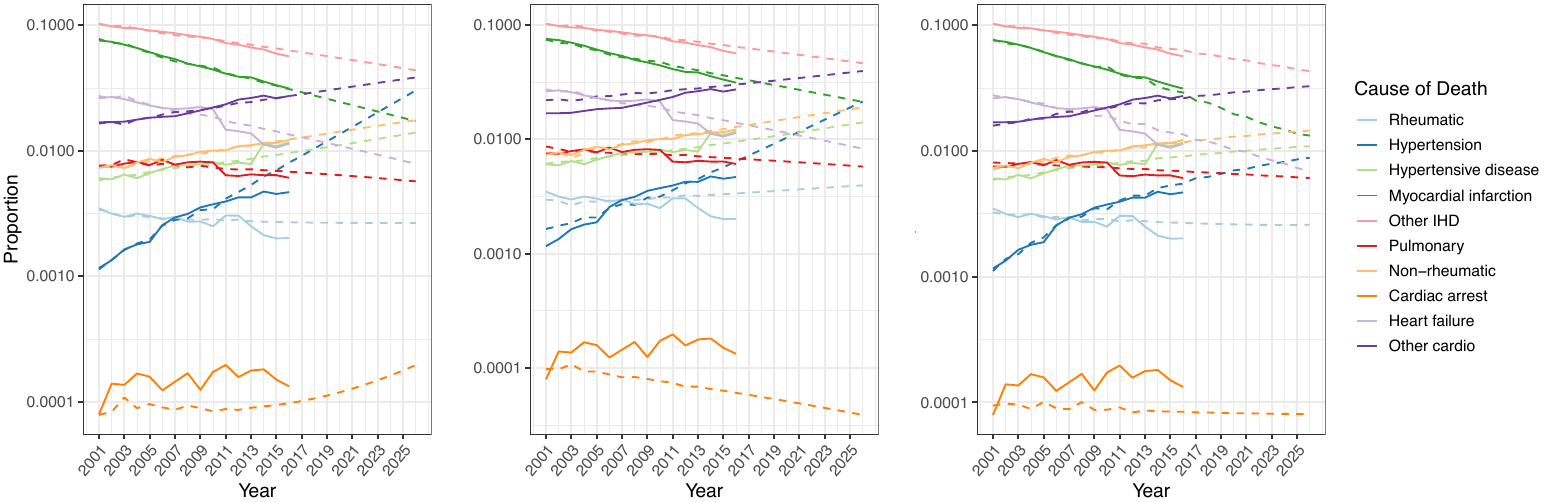}\label{fig:all_proj_f}}
\quad
\caption{Forecast of cause-specific mortality up to 2026 in our application to England and Wales data from the HCD database, disaggregated for cardiovascular causes of death. Solid lines represent the observed mortality by cause proportions, and dashed lines show the forecast using the CLR, ILR (with zeros removed), and $\alpha$-transformations (L--R). Mortality by cause is shown for males (top row) and females (bottom row). This figure omits non-cardiovascular causes for presentation purposes.} \label{fig:cardio_results}
\end{figure}

Results in Figure~\ref{fig:cardio_results} are consistent with the broader observations that overall mortality experienced due to cardiovascular causes in the UK has been improving since the 1960s \citep{BHF23, NHS22, ONS21}, although forecasts suggest that an expected decline in the major cardiovascular causes (myocardial infarction and pulmonary heart disease) will be offset by forecast increases in the ``other heart" cause category. Again, results from the $\alpha$-transformation follow the observed data over time more closely compared to CLR and ILR with zeros removed. Moreover, the standard LRA approaches, where a value of 0.25 or 0.5 was added to the zeros, tended to forecast higher proportions for causes with the lowest proportion of deaths (in this case, cardiac arrest), which is offset by lower forecast proportions across all other causes (results shown in Appendix \ref{sec:replacementzeros}). This result is consistent with the fact that these approaches arbitrarily introduce small death counts where there are none. 

\subsection{Results: US data}\label{sec:cardio_results_US} 

\begin{figure}[!htb]

{\includegraphics[width=14.5cm]{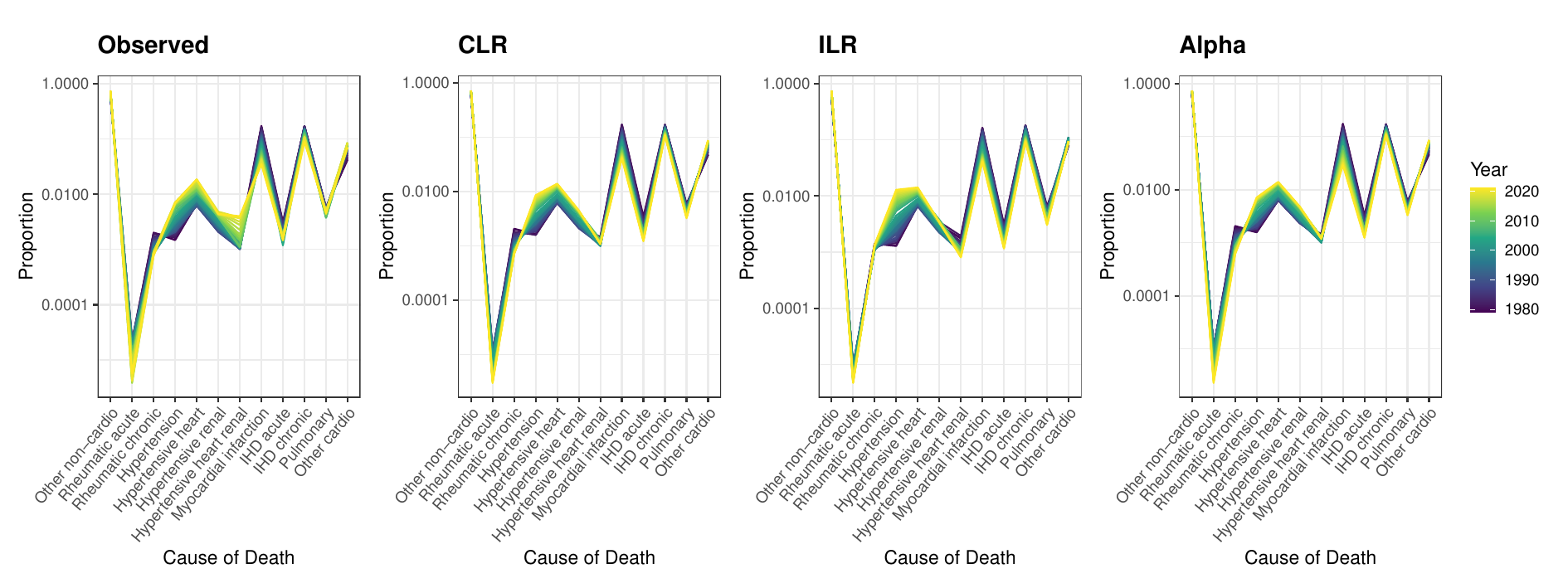}\label{fig:us_cardio_compare_m_0.7}}
\quad
{\includegraphics[width=15.5cm]{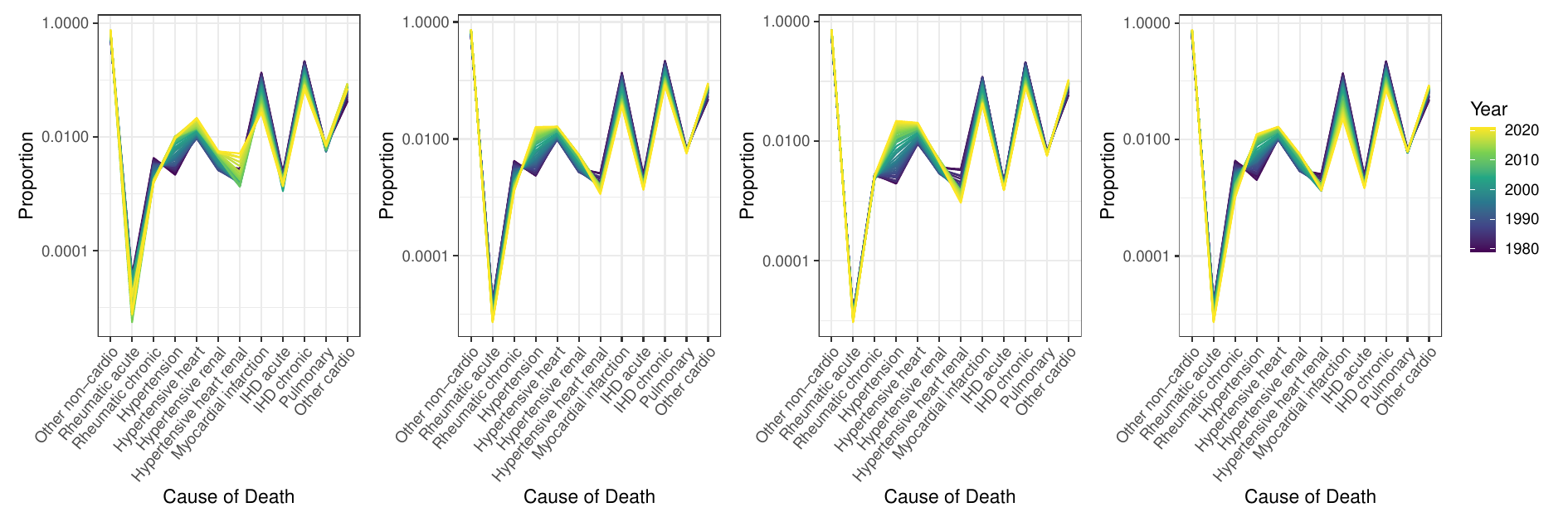}\label{fig:us_cardio_compare_f_0.9}}
\quad
\caption{Male (top row) and female (bottom row) mortality by cause in our application to US data from \cite{HMD24}, disaggregated for cardiovascular causes of death. The figures show the movement in actual proportion of deaths for each cause from 1979 to 2021 (left column), while the remaining three columns present results from applying CLR, ILR (with zeros removed) and $\alpha$-transformations, respectively.} \label{fig:us_cardio_compare}
\end{figure}

For the larger US cause-of-death dataset, Figure~\ref{fig:us_cardio_compare} shows the movement in actual proportion of deaths for each cause over the historical data for US death counts, in a similar way to Figure~\ref{fig:cardio_compare}. 

\begin{figure}[!htb]
{\includegraphics[width=1.15\textwidth]{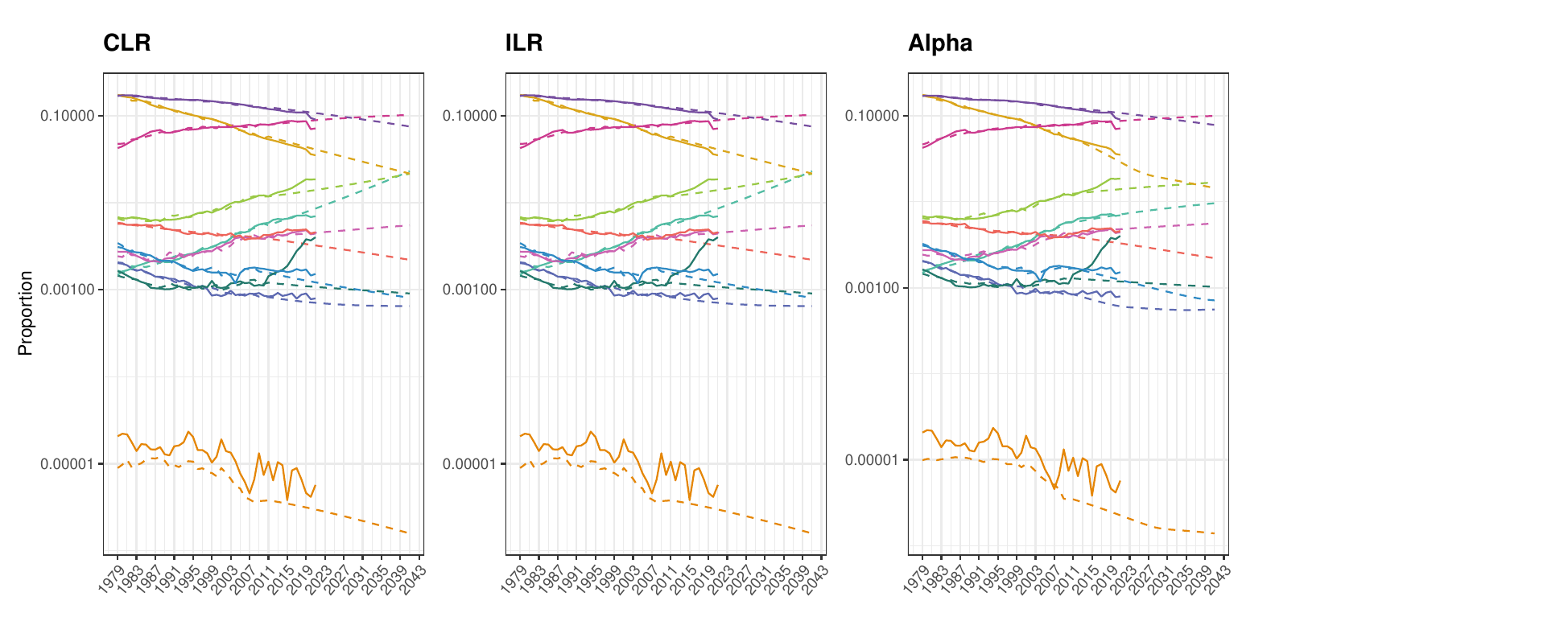}\label{fig:us_all_proj_m}}
\\
{\includegraphics[width=1.1\textwidth]{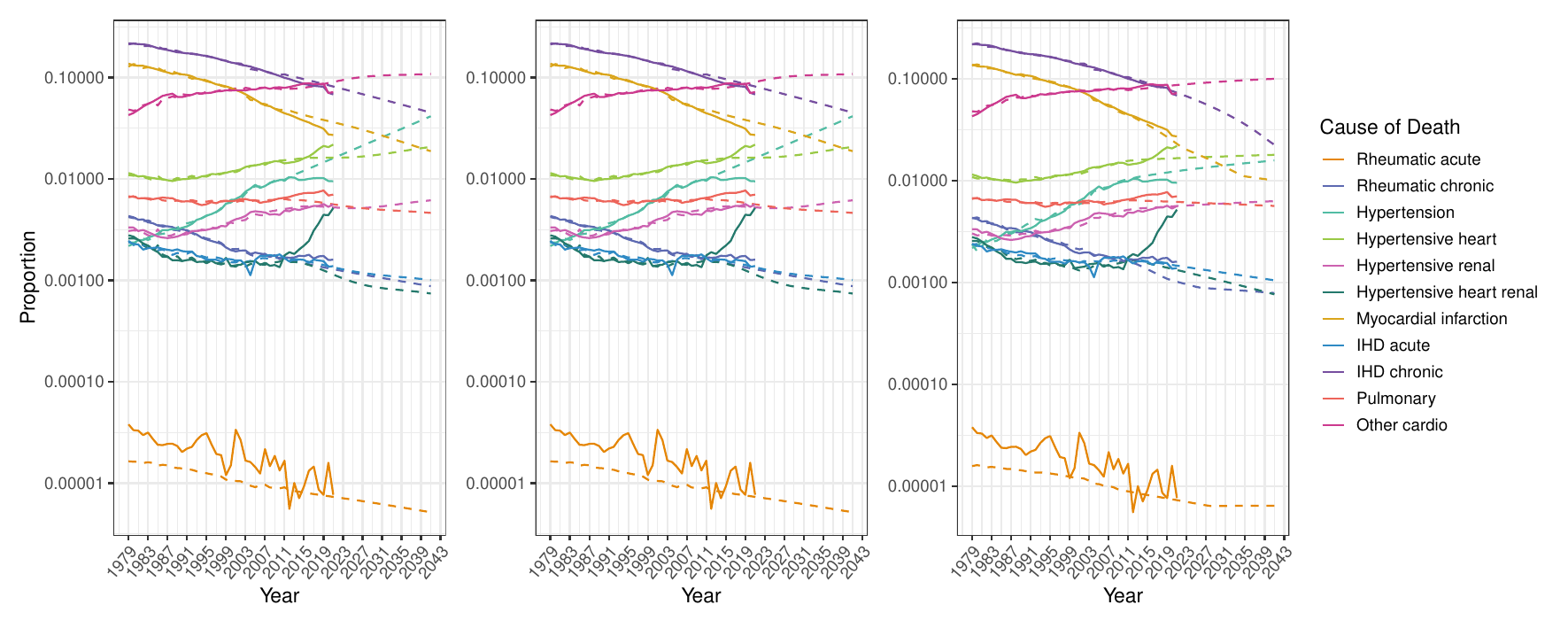}\label{fig:us_all_proj_f}}
\quad
\caption{Forecast of cause-specific mortality up to 2051 in our application to US data from \cite{HMD24}, disaggregated for cardiovascular causes of death. Solid lines represent the observed mortality by cause proportions, and dashed lines show the forecast using the CLR, ILR (with zeros removed), and $\alpha$-transformations (L--R). Mortality by cause is shown for males (top row) and females (bottom row). This figure omits non-cardiovascular causes for presentation purposes.} \label{fig:us_cardio_results}
\end{figure}

The $\alpha$-transformation results followed the observed data over time more closely compared to CLR and ILR with zeros removed. This is shown in Table~\ref{tab:cardio_results_US} and Figure~\ref{fig:us_cardio_results}. Moreover, the standard LRA approaches, where a value of~0.25 or~0.5 was added to the zeros, tended to forecast higher proportions for causes with the lowest proportion of deaths (in this case, cardiac arrest), which is offset by lower forecast proportions across all other causes (results shown in Appendix \ref{sec:replacementzeros}). This result was consistent with the arbitrary introduction of a small death count where none existed. More importantly, compared with England and Wales data, the forecast performance using the $\alpha$-transformation was even further improved in the application. Indeed, it suggests that, 
with a larger volume of data available, our proposed approach can exhibit greater forecasting performance compared to existing log-ratio transformation approaches.

\begin{table}[!htb] 
\caption{Forecast performance on test data, applying CLR, ILR, and the $\alpha$-transformation coupled with the LC model to US data from \cite{HMD24}, disaggregated for cardiovascular causes of death. For each metric and gender, the bolded values correspond to the error using optimal values of $\alpha$ tuned based on cross-validation. In contrast, underlined values correspond to the lowest metric in the out-of-sample forecast.}
\label{tab:cardio_results_US} 
\centering
\tabcolsep 0.27in
\begin{tabular}{@{}lllll@{}} 
 \toprule[1.5pt]
Method & \multicolumn{2}{c}{RMSE $\times$ 100} & \multicolumn{2}{c}{MAE $\times$ 100}  \\ [0.5ex] 
        & Male & Female & Male & Female \\
\cmidrule{1-5}
CLR (zeros omitted) & 0.3370 & 0.3819 & 0.1417 & 0.1650\\
CLR (0.25 zero replacement) & 0.3566 & 0.4393 & 0.1541 & 0.2049\\
CLR (0.5 zero replacement) & 0.3477 & 0.4278 & 0.1467 & 0.1947\\
ILR (zeros omitted) & 0.3370 & 0.3819 & 0.1417 & 0.1650\\
ILR (0.25 zero replacement) & 0.3566 & 0.4393 & 0.1541 & 0.2049\\
ILR (0.5 zero replacement) & 0.3477 & 0.4278 & 0.1467 & 0.1947\\
$\alpha$ = 0.3 & 0.3072 & 0.3138 & 0.1314 & 0.1439\\ 
$\alpha$ = 0.5 & 0.2905 & 0.2777 & \underline{0.1268} & 0.1300\\ 
$\alpha$ = 0.7 & \underline{\bf{0.2877}} & 0.2518 & \bf{0.1299} & \underline{0.1202}\\
$\alpha$ = 0.9 & 0.3095 & \underline{\bf{0.2516}} & 0.1355 & \bf{0.1238}\\
$\alpha$ = 1 (RDA) & 0.3435 & 0.2691 & 0.1414 & 0.1287\\ 
\bottomrule[1.5pt]
\end{tabular}
\end{table}

In summary, the point forecast results across both applications suggested that the $\alpha$-transformation, a generalisation of the log-ratio transformation to a broader class of transformations, was an effective way to address zero counts in compositional data, especially compared to ad-hoc methods of adding small death counts. In Appendix \ref{app:sensitivity_alpha}, we performed a sensitivity analysis to assess how much the performance in the two applications depended on the precise $\alpha$ value chosen. Overall, results showed that forecasting performance was largely unaffected when the value of $\alpha$ changed within the tolerance of 0.1 that we employed when tuning this parameter in Section \ref{sec:alpha-optimisation}. 

\subsection{Interval forecasts}\label{sec:interval_forecasts} 


To further understand the projected deaths using the $\alpha$-transformation, we used interval forecasts to quantify the uncertainty around the point forecast and a further source of (probabilistic) comparison between different methods across both applications of the HCD data (i.e. England and Wales and US death counts). Briefly, the interval forecasts were produced by adapting the proposed method of \citet{SH20} for use with the CLR, ILR, and $\alpha$-transformations, and involved the following steps.
\begin{enumerate}
\item Transform the compositional data into the real space using the three methods explored (CLR, ILR, and the proposed $\alpha$-transformation). Construct the point forecast as per Sections~\ref{sec:cardio_results} and~\ref{sec:cardio_results_US}.
\item Bootstrap (sample with replacement) the forecast component scores (i.e., $b_{u, c}$ or the age- and cause-specific coefficients which vary over time) and the model fit errors (i.e., $\epsilon_{t, u, c}$) in equation~\eqref{eqn:LC CODA CLR}. By doing this a large number of times and then taking the empirical quantiles (here, 90\% intervals are shown), upper and lower bounds for the interval forecast in real space is produced.
\item Transform the interval forecast from the real space to the simplex for inference using the corresponding inverse CLR, ILR, or $\alpha$-transformations. Finally, add back the geometric mean as per the original point estimate approach discussed per equation~\eqref{eqn:LC CODA CLR}.
\end{enumerate}

\begin{figure}[!htb]
{\includegraphics[width=13.5cm]{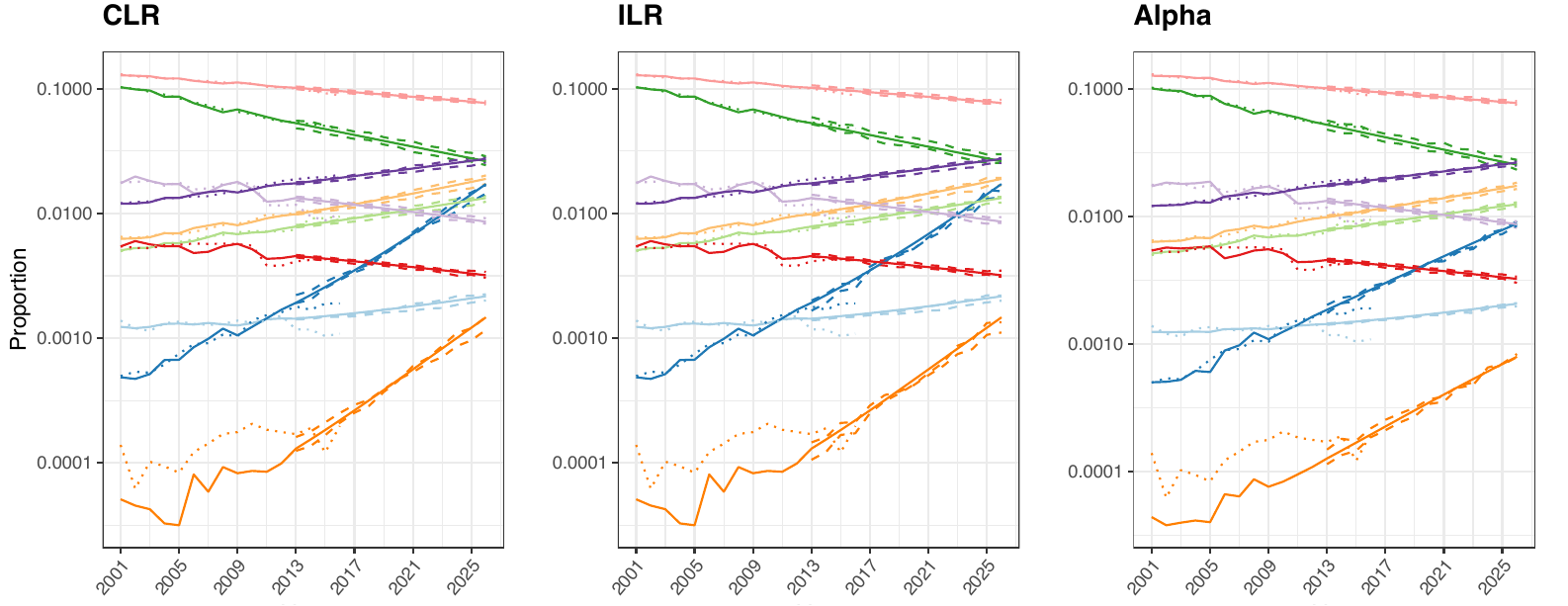}\label{fig:uk_interval_m_0.1}}
\quad
{\includegraphics[width=15.5cm]{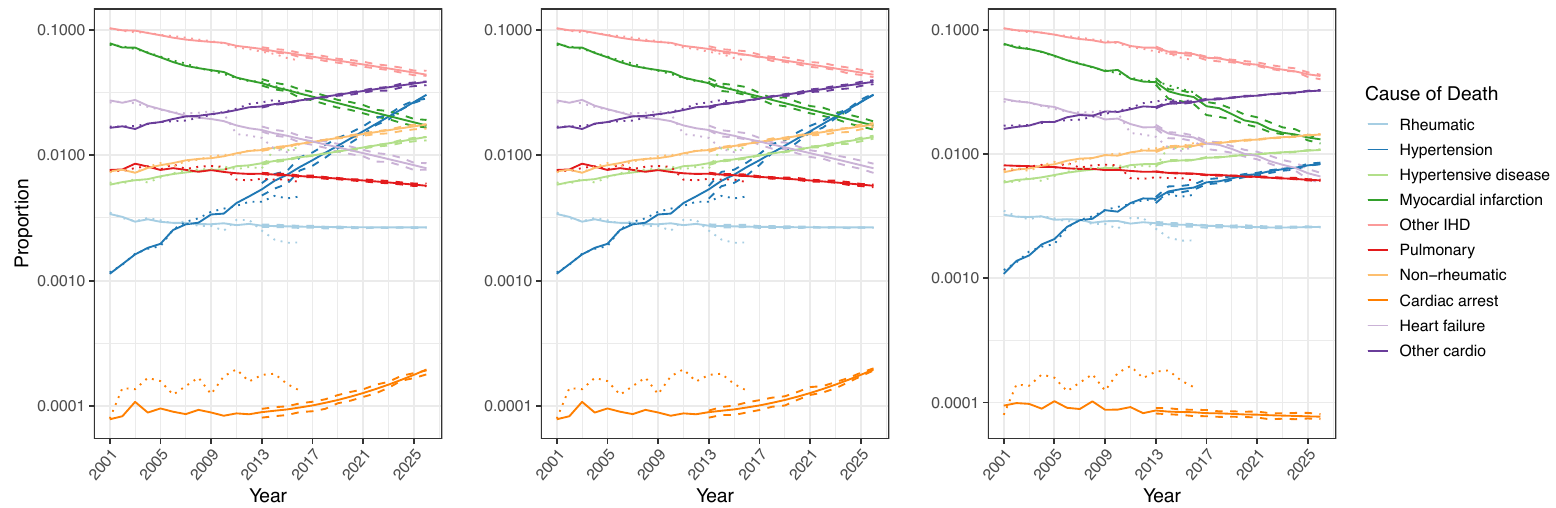}\label{fig:uk_interval_f_0.8}}
\quad
\caption{Male (top row) and female (bottom row) 90\% interval forecasts up to 2026 in our application to England and Wales data from \cite{HMD24}, disaggregated for cardiovascular causes of death.}\label{fig:uk_interval}
\end{figure}

\begin{figure}[!htb]
{\includegraphics[width=13.1cm]{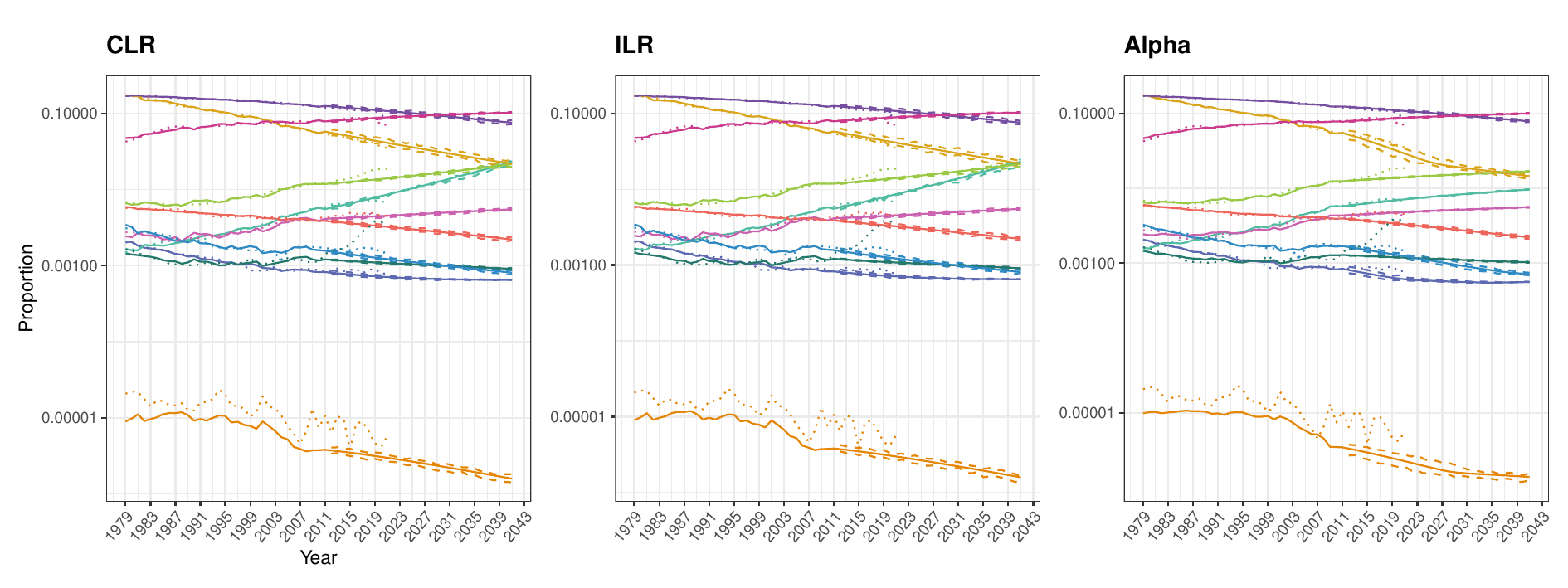}\label{fig:us_interval_m_0.7}}
\quad
{\includegraphics[width=15.7cm]{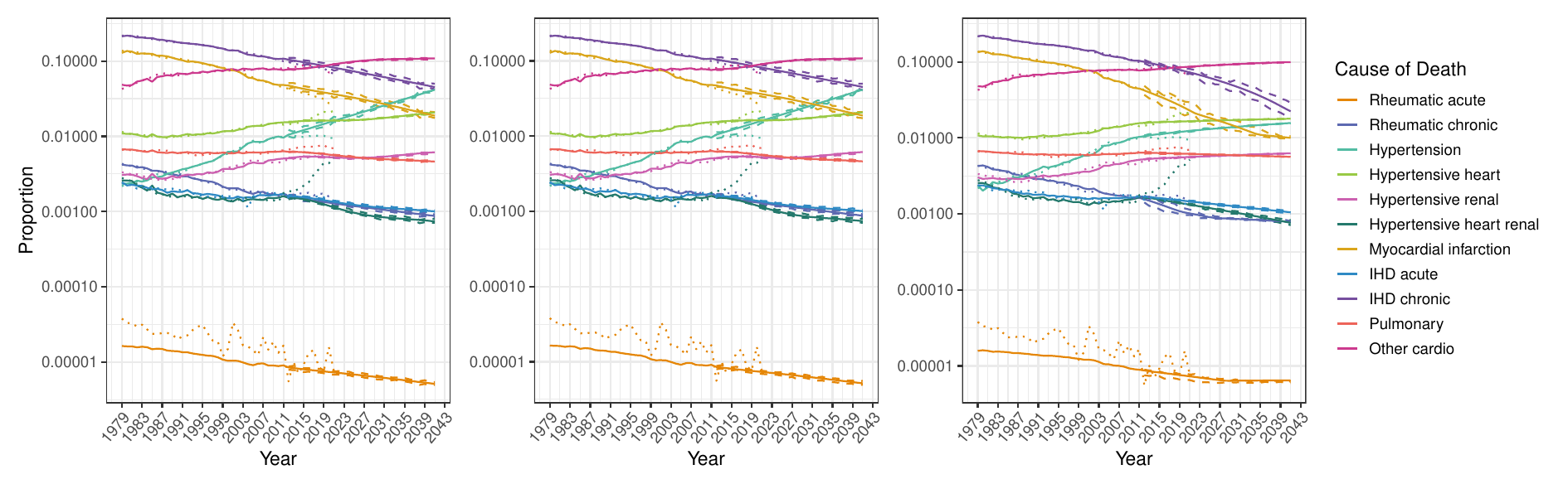}\label{fig:us_interval_f_0.9}}
\quad
\caption{Male (top row) and female (bottom row) 90\% interval forecasts up to 2051 in our application to US data from \cite{HMD24}, disaggregated for cardiovascular causes of death.} \label{fig:us_interval}
\end{figure}

Results for the interval forecasts for both applications are presented in Figures~\ref{fig:uk_interval} and ~\ref{fig:us_interval}, where the $\alpha$ parameters used in producing interval forecasts were optimised via the interval score approach of \citet{SH20}. Overall, the results across CLR, ILR, and the $\alpha$-transformation were largely consistent with 
the corresponding point forecasts results shown previously in Figures~\ref{fig:cardio_results} and ~\ref{fig:us_cardio_results}. Nevertheless, the interval forecast offers an additional view of uncertainty around the point forecast, and reflects the possible extents to which the composition of mortality across different causes could change into the future based on each model.

\subsection{Alternative approaches and future directions}\label{sec:alternatives}

In this section, we consider an alternative approach to the $\alpha$-transformation for forecasting mortality by cause. Specifically, we consider the multinomial logistic regression (MLR) of \citet{AAS15}, and compare its forecast performance to the $\alpha$-transformation. 

The MLR model is often used to detect factors significantly influencing a response with several competing outcomes. In the literature, numerous applications of the MLR model have been undertaken in cause-of-death analysis over the past three decades. For example, \cite{ENH90} used eight categorical and continuous independent variables, including marital status, education, and birth weight, to model five infant cause-specific mortality rates. \cite{LWC06} applied MLR to model the distribution of neonatal deaths in countries with poor data \citep[see][for related work]{JLF+10}. \cite{SBL+13} employed MLR to redistribute unknown or ill-defined deaths, while \cite{PCL06} used it as to account for the impact of the tenth revision of the International Classification of Diseases (ICD). 

\begin{figure}[!htb]
\centering
{\includegraphics[width=15.5cm]{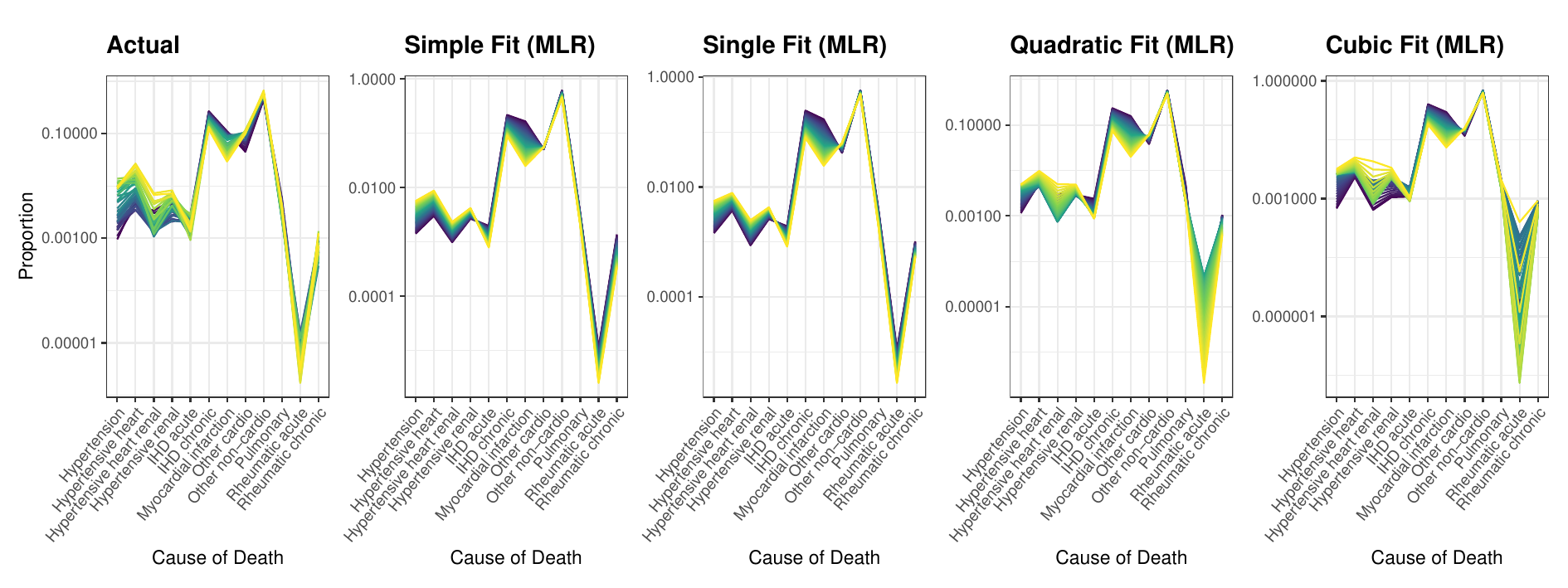}\label{fig:us_MLR_comparison_m}}
\quad
\caption{Male mortality by cause using US data from \cite{HMD24}, disaggregated for cardiovascular causes of death. The figures show the movement in actual proportion of deaths for each cause from 1979 to 2021 (left column), while the remaining four columns present results from applying MLR simple, single, quadratic, and cubic regressions, respectively.}\label{fig:us_MLR_cardio_compare}
\end{figure}

\begin{figure}[!htb]
\centering
{\includegraphics[width=0.95\textwidth]{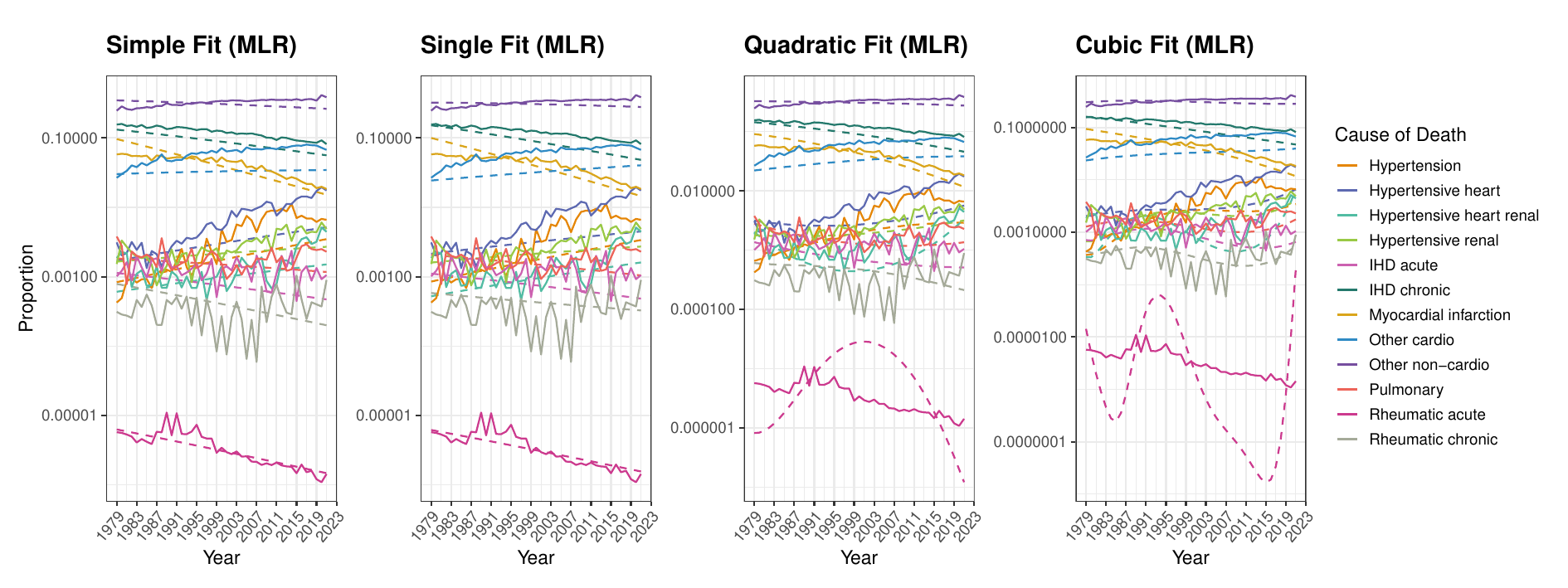}\label{fig:us_MLR_all_proj_m}}
\quad
\caption{Fits of cause-specific male mortality in our application to US data from \cite{HMD24}, disaggregated for cardiovascular causes of death. Solid lines represent the observed mortality by cause proportions, and dashed lines show the fit using MLR regressions. This figure omits non-cardiovascular causes for presentation purposes.} \label{fig:us_MLR_cardio_results}
\end{figure}

For illustrative purposes, we applied the MLR model to US male cause-of-death counts only,  disaggregated for cardiovascular causes as per the application in Section~\ref{sec:cardio_results_US}. 
The forecast performance from applying MLR was assessed using the sum of the squared residual errors. Based on this, we found that the single and simple MLR performed best when compared against the quadratic and cubic MLR. We present results for these in Figures \ref{fig:us_MLR_cardio_compare} and \ref{fig:us_MLR_cardio_results}, which are analogous to those presented earlier in Figure~\ref{fig:us_cardio_compare} and Figure~\ref{fig:us_cardio_results}. Note in assessing the fits, the problem of zeros was still present in the actual death rates by cause; we handled this by adding a~0.01 death count before calculating mortality rates and taking logarithms.

To compare with the forecast performance using CODA methods and shown in Table~\ref{tab:cardio_results_US}, we calculated the equivalent RMSE and MAE (scaled by 100) for the MLR application to US male death counts. In this application, the simple MLR produced RMSE and MAE of 2.289 and 1.126, whereas the single MLR produced RMSE and MAE of 2.040 and 1.038. This is substantially higher than the errors of 0.2877 and 0.1299 when we apply the CODA method using an $\alpha$-transformation. We conjecture similar results would also arise for the case of the US female cause-of-death count data, as well as the England and Wales data. Overall, the comparison indicates that, perhaps not surprisingly, CODA approaches perform better when forecasting using compositional data.


Beyond the MLR model, another method to address the problem of zeros in composition data is applying the Dirichlet distribution. This idea has previously been explored by \citet{TS18} and \cite{GN23} in modifying the log-likelihood of the Dirichlet distribution. Such approaches have been applied in other fields, including biology and chromosome detection \citet{TWY+22}. Further exploration of the Dirichlet composition distribution in understanding mortality by cause would further the understanding of mortality forecasting by cause. Finally, a forecast reconciliation approach can be adopted to ensure forecast coherence instead of applying compositional data analysis \citep{LLL+19}. Such approaches address the potential problems arising when sub-population mortality forecasts do not sum up to the aggregate forecast, and they could be considered an alternative approach where there are few zeros in subgroups.

\section{Conclusion}\label{sec:discussion}

In this paper, we have introduced the $\alpha$-transformation, coupled with a Lee-Carter model for mortality modelling, as a statistical method to handle cause-of-death compositional data with zero values. Using an expanding window cross-validation approach to select $\alpha$, we presented two applications to death counts by cause, disaggregated for cardiovascular causes on England and Wales data from 2001 to 2016, and on US data from 1979 to 2021. Forecasts using the $\alpha$-transformation tend to perform better than those produced using standard log-ratio transformations and is particularly evident in the application to US death counts by cause, having more years of historical data. 

We tested a single model (LC) in the compositional framework and focused on heart-related causes of death, where the data set includes zero counts for several years and age bands. Mortality forecasting by cause may be further enhanced by combining the $\alpha$-transformation with variations of the LC model, for example, a model which decomposes cause-specific variation into joint and individual variation \citep{KEK+19}, or using non-parametric techniques such as smoothers or tensor decompositions \citep{ZHH23}. Also, rather than adding a small death count or removing zeros entirely, other approaches could be compared against the $\alpha$-transformation, including ``borrowing" from a neighbouring age (for the same cause) or smoothing over similar causes (for the same age), along with other imputation methods \citep{LFT21}. We leave such investigations as avenues for future study.

One feature of the death counts by cause for both England and Wales and the US, which is true of many other cause-of-death datasets in other countries, is that zero counts of death for multiple causes tend to occur across consecutive years and/or adjacent age groups. In other settings with fewer or no zeros count, and where the occurrence of the zeros is more sporadic, simpler approaches, such as adding a small value to enable LRA may have fewer implications on the analysis and conclusions relative to using the $\alpha$-transformation \citep{TS22}. Conversely, for older ages and emerging causes with only recent data (including COVID-19), reflecting true zeros in the data in a statistically more rigorous and data-driven manner, as the $\alpha$-transformation does, is expected to produce more accurate forecasts.

Whilst CODA is useful in capturing dependencies between causes arising due to the compositional nature of the data, other dependencies, such as co-morbidities, can arise irrespective of how the data are treated. An important avenue of future research is how methods such as $\alpha$-transformation could be coupled with techniques that can account for such dependencies. Indeed, an essential application of CODA for life insurers is to enhance the understanding of morbidity and mortality risks. CODA can also be used to investigate the risk implications across different subgroups of insured lives and exposures, and we anticipate the $\alpha$-transformation will play a useful role in modelling compositional data arising from these other settings. Finally, the results from the application suggest that while the aggregate cardiovascular death counts are expected to reduce, some granular causes of death within the cardiovascular cause are expected to increase, particularly for males across England and Wales. Analysis using US death counts indicate slight decreases across all granular cardiovascular causes. These findings should be further investigated, along with other causes. 

\section*{Acknowledgments}

The second author was supported by an Australian Research Council Discovery Project, DP230102250, and the third author was supported by an Australian Research Council Discovery Project, DP230101908. The authors are grateful for the comments by the participants at the Insurance Data Science conference in 2023 and the Conference in Celebration of David Wilkie's 90th birthday in 2024.

\section*{Data Availability Statement}

The data and code that support the findings of this study are openly available at \url{https://github.com/zm-dong/coda_cause_mortality}.

\newpage
\bibliographystyle{apalike}
\bibliography{CoDA.bib}

\begin{thebibliography}{}

\bibitem[Aitchison, 1982]{A82}
Aitchison, J. (1982).
\newblock The statistical analysis of compositional data.
\newblock {\em Journal of the Royal Statistical Society: Series B
  (Methodological)}, 44(2):139--160.

\bibitem[Alai et~al., 2015]{AAS15}
Alai, D.~H., Arnold, S., and Sherris, M. (2015).
\newblock Modelling cause-of-death mortality and the impact of
  cause-elimination.
\newblock {\em Annals of Actuarial Science}, 9(1):167--186.

\bibitem[Arnold and Sherris, 2013]{AS13}
Arnold, S. and Sherris, M. (2013).
\newblock Forecasting mortality trends allowing for cause-of-death mortality
  dependence.
\newblock {\em North American Actuarial Journal}, 17(4):273--282.

\bibitem[{Basel Committee on Banking Supervision}, 2013]{B13}
{Basel Committee on Banking Supervision} (2013).
\newblock Longevity risk transfer markets: market structure, growth drivers and
  impediments, and potential risks.
\newblock URL: \url{https://www.bis.org/publ/joint34.pdf}.

\bibitem[Bergeron-Boucher et~al., 2017]{BCO17}
Bergeron-Boucher, M.-P., Canudas-Romo, V., Oeppen, J., and Vaupel, J.~W.
  (2017).
\newblock Coherent forecasts of mortality with compositional data analysis.
\newblock {\em Demographic Research}, 37:527--566.

\bibitem[Bergeron-Boucher and Kj{\ae}rgaard, 2022]{BK22}
Bergeron-Boucher, M.-P. and Kj{\ae}rgaard, S. (2022).
\newblock Mortality forecasting at age 65 and above: {A}n age-specific
  evaluation of the {Lee-Carter} model.
\newblock {\em Scandinavian Actuarial Journal}, 2022(1):64--79.

\bibitem[Bergeron-Boucher et~al., 2022]{BSS+22}
Bergeron-Boucher, M.-P., Strozza, C., Simonacci, V., and Oeppen, J. (2022).
\newblock Modeling and forecasting healthy life expectancy with compositional
  data analysis.
\newblock {\em SocArXiv. July}, 9.

\bibitem[Booth et~al., 2002]{BMS02}
Booth, H., Maindonald, J., and Smith, L. (2002).
\newblock Applying {Lee-Carter} under conditions of variable mortality decline.
\newblock {\em Population Studies}, 56(3):325--336.

\bibitem[Booth et~al., 2005]{BTS05}
Booth, H., Tickle, L., Smith, L., et~al. (2005).
\newblock Evaluation of the variants of the {Lee-Carter} method of forecasting
  mortality: {A} multi-country comparison.
\newblock {\em New Zealand Population Review}, 31(1):13--34.

\bibitem[Box and Cox, 1964]{BC64}
Box, G.~E. and Cox, D.~R. (1964).
\newblock An analysis of transformations.
\newblock {\em Journal of the Royal Statistical Society: Series B
  (Methodological)}, 26(2):211--243.

\bibitem[{British Heart Foundation}, 2023]{BHF23}
{British Heart Foundation} (2023).
\newblock {UK Factsheet}.
\newblock URL:
  \url{https://www.bhf.org.uk/-/media/files/for-professionals/research/heart-statistics/bhf-cvd-statistics-uk-factsheet.pdf}.

\bibitem[Cairns et~al., 2006]{CBD06}
Cairns, A.~J., Blake, D., and Dowd, K. (2006).
\newblock A two-factor model for stochastic mortality with parameter
  uncertainty: {T}heory and calibration.
\newblock {\em Journal of Risk and Insurance}, 73(4):687--718.

\bibitem[Chai and Draxler, 2014]{CD14}
Chai, T. and Draxler, R.~R. (2014).
\newblock {Root mean square error {(RMSE)} or mean absolute error
  {(MAE)}?--Arguments against avoiding RMSE in the literature}.
\newblock {\em Geoscientific Model Development}, 7(3):1247--1250.

\bibitem[Eberstein et~al., 1990]{ENH90}
Eberstein, I.~W., Nam, C.~B., and Hummer, R.~A. (1990).
\newblock {Infant mortality by cause of death: Main and interaction effects}.
\newblock {\em Demography}, 27:413--430.

\bibitem[Gao and Shi, 2021]{GS21}
Gao, G. and Shi, Y. (2021).
\newblock Age-coherent extensions of the {Lee--Carter} model.
\newblock {\em Scandinavian Actuarial Journal}, 2021(10):998--1016.

\bibitem[Graziani and Nigri, 2023]{GN23}
Graziani, R. and Nigri, A. (2023).
\newblock An age–period–cohort model in a dirichlet framework: {A} coherent
  causes of death estimation.
\newblock Technical report, SocArXiv.
\newblock URL: \url{https://ideas.repec.org/p/osf/socarx/856yw.html}.

\bibitem[Greenacre, 2021]{G21}
Greenacre, M. (2021).
\newblock Compositional data analysis.
\newblock {\em Annual Review of Statistics and its Application}, 8:271--299.

\bibitem[Greenacre, 2024]{G23}
Greenacre, M. (2024).
\newblock The chipower transformation: a valid alternative to logratio
  transformations in compositional data analysis.
\newblock Working paper, arXiv.

\bibitem[Greenacre and Grunsky, 2019]{GG19}
Greenacre, M. and Grunsky, E. (2019).
\newblock {The isometric logratio transformation in compositional data
  analysis: A practical evaluation}.
\newblock Working paper 1627, Universitat Pompeu Fabra.

\bibitem[Grifoll et~al., 2019]{GOE19}
Grifoll, M., Ortego, M., and Egozcue, J. (2019).
\newblock Compositional data techniques for the analysis of the container
  traffic share in a multi-port region.
\newblock {\em European Transport Research Review}, 11(1):1--15.

\bibitem[Hodson, 2022]{H22}
Hodson, T.~O. (2022).
\newblock {Root-mean-square error {(RMSE)} or mean absolute error {(MAE)}: When
  to use them or not}.
\newblock {\em Geoscientific Model Development}, 15(14):5481--5487.

\bibitem[Holford, 1983]{H83}
Holford, T.~R. (1983).
\newblock The estimation of age, period and cohort effects for vital rates.
\newblock {\em Biometrics}, pages 311--324.

\bibitem[{Human Cause-of-death Data series}, 2024]{HMD24}
{Human Cause-of-death Data series} (2024).
\newblock {French Institute for Demographic Studies (France), Max Planck
  Institute for Demographic Research (Germany) and the University of
  California, Berkeley (USA).}

\bibitem[Hyndman et~al., 2013]{HBY13}
Hyndman, R.~J., Booth, H., and Yasmeen, F. (2013).
\newblock Coherent mortality forecasting: the product-ratio method with
  functional time series models.
\newblock {\em Demography}, 50(1):261--283.

\bibitem[Johnson et~al., 2010]{JLF+10}
Johnson, H.~L., Liu, L., {Fischer-Walker}, C., and Black, R.~E. (2010).
\newblock Estimating the distribution of causes of death among children age
  1-59 months in high-mortality countries with incomplete death certification.
\newblock {\em International Journal of Epidemiology}, 39:1103--1114.

\bibitem[Kjaergaard et~al., 2020]{KEB+20}
Kjaergaard, S., Ergemen, Y.~E., Bergeron-Boucher, M.-P., Oeppen, J., and
  Kallestrup-Lamb, M. (2020).
\newblock Longevity forecasting by socio-economic groups using compositional
  data analysis.
\newblock {\em Journal of the Royal Statistical Society: Series A (Statistics
  in Society)}, 183(3):1167--1187.

\bibitem[Kjaergaard et~al., 2019]{KEK+19}
Kjaergaard, S., Ergemen, Y.~E., {Kallestrup-Lamb}, M., Oeppen, J., and
  {Lindahl-Jacobsen}, R. (2019).
\newblock Forecasting causes of death using compositional data analysis: the
  case of cancer deaths.
\newblock {\em Journal of the Royal Statistical Society Series C: Applied
  Statistics}, 68(5):1351--1370.

\bibitem[Lawn et~al., 2006]{LWC06}
Lawn, J.~E., {Wilczynska-Ketende}, K., and Cousens, S.~N. (2006).
\newblock Estimating the causes of 4 million neonatal deaths in the year 2000.
\newblock {\em International Journal of Epidemiology}, 35:706--718.

\bibitem[Lee and Miller, 2001]{LM01}
Lee, R. and Miller, T. (2001).
\newblock Evaluating the performance of the lee-carter method for forecasting
  mortality.
\newblock {\em Demography}, 38:537--549.

\bibitem[Lee and Carter, 1992]{LC92}
Lee, R.~D. and Carter, L.~R. (1992).
\newblock Modeling and forecasting {US} mortality.
\newblock {\em Journal of the American Statistical Association: Applications \&
  Case Studies}, 87(419):659--671.

\bibitem[Li et~al., 2019a]{LLL19}
Li, H., Li, H., Lu, Y., and Panagiotelis, A. (2019a).
\newblock A forecast reconciliation approach to cause-of-death mortality
  modeling.
\newblock {\em Insurance: Mathematics and Economics}, 86:122--133.

\bibitem[Li et~al., 2019b]{LLL+19}
Li, H., Li, H., Lu, Y., and Panagiotelis, A. (2019b).
\newblock A forecast reconciliation approach to cause-of-death mortality
  modeling.
\newblock {\em Insurance: Mathematics and Economics}, 86:122--133.

\bibitem[Li and Lu, 2019]{LL19}
Li, H. and Lu, Y. (2019).
\newblock Modeling cause-of-death mortality using hierarchical archimedean
  copula.
\newblock {\em Scandinavian Actuarial Journal}, 2019(3):247--272.

\bibitem[Li and Lee, 2005]{LL05}
Li, N. and Lee, R. (2005).
\newblock Coherent mortality forecasts for a group of populations: An extension
  of the lee-carter method.
\newblock {\em Demography}, 42:575--594.

\bibitem[Lubbe et~al., 2021]{LFT21}
Lubbe, S., Filzmoser, P., and Templ, M. (2021).
\newblock Comparison of zero replacement strategies for compositional data with
  large numbers of zeros.
\newblock {\em Chemometrics and Intelligent Laboratory Systems}, 210:104248.

\bibitem[Martin-Fernandez et~al., 2003]{MBP03}
Martin-Fernandez, J.~A., Barcelo-Vidal, C., and Pawlowsky-Glahn, V. (2003).
\newblock Dealing with zeros and missing values in compositional data sets
  using nonparametric imputation.
\newblock {\em Mathematical Geology}, 35:253--278.

\bibitem[{National Institute for Health and Care Excellence}, 2023]{CKS23}
{National Institute for Health and Care Excellence} (2023).
\newblock {CVD} risk assessment and management: What is the impact of {CVD}?
\newblock URL:
  \url{https://cks.nice.org.uk/topics/cvd-risk-assessment-management/background-information/burden-of-cvd/}.

\bibitem[{NHS}, 2023]{NHS22}
{NHS} (2023).
\newblock Cardiovascular disease.
\newblock URL: \url{https://www.nhs.uk/conditions/cardiovascular-disease/}.

\bibitem[Oeppen, 2008]{O08}
Oeppen, J. (2008).
\newblock Coherent forecasting of multiple-decrement life tables: a test using
  {Japanese} cause of death data.

\bibitem[{Office for National Statistics}, 2021]{ONS21}
{Office for National Statistics} (2021).
\newblock Ischaemic heart diseases deaths including comorbidities, england and
  wales: 2019 registrations.
\newblock URL:
  \url{https://www.ons.gov.uk/peoplepopulationandcommunity/birthsdeathsandmarriages/deaths/bulletins/ischaemicheartdiseasesdeathsincludingcomorbiditiesenglandandwales/2019registrations}.

\bibitem[Park et~al., 2006]{PCL06}
Park, Y., Choi, J.~W., and Lee, D.-H. (2006).
\newblock {A parametric approach for measuring the effect of the 10th revision
  of the international classification of diseases}.
\newblock {\em Journal of the Royal Statistical Society: Series C},
  55(5):677--697.

\bibitem[Racine, 2000]{R00}
Racine, J. (2000).
\newblock Consistent cross-validatory model-selection for dependent data:
  hv-block cross-validation.
\newblock {\em Journal of econometrics}, 99(1):39--61.

\bibitem[Raleigh et~al., 2022]{RJW22}
Raleigh, V., Jefferies, D., and Wellings, D. (2022).
\newblock Cardiovascular diseases in {England}: supporting leaders to take
  action.
\newblock URL:
  \url{https://www.kingsfund.org.uk/publications/cardiovascular-disease-england}.

\bibitem[Renshaw and Haberman, 2003]{RH03}
Renshaw, A.~E. and Haberman, S. (2003).
\newblock {Lee--Carter mortality forecasting with age-specific enhancement}.
\newblock {\em Insurance: Mathematics and Economics}, 33:255--272.

\bibitem[Renshaw and Haberman, 2006]{RH06}
Renshaw, A.~E. and Haberman, S. (2006).
\newblock A cohort-based extension to the {Lee}-{Carter} model for mortality
  reduction factors.
\newblock {\em Insurance: Mathematics and economics}, 38(3):556--570.

\bibitem[Schnaubelt, 2019]{S19}
Schnaubelt, M. (2019).
\newblock A comparison of machine learning model validation schemes for
  non-stationary time series data.
\newblock Technical report, FAU Discussion Papers in Economics.

\bibitem[Shahraz et~al., 2013]{SBL+13}
Shahraz, S., Bhalla, K., Lozano, R., Bartels, D., and Murray, C. J.~L. (2013).
\newblock {Improving the quality of road injury statistics by using regression
  models to redistribute ill-defined events}.
\newblock {\em Injury Prevention}, 19(1):1--5.

\bibitem[Shang and Haberman, 2020]{SH20}
Shang, H.~L. and Haberman, S. (2020).
\newblock Forecasting age distribution of death counts: {A}n application to
  annuity pricing.
\newblock {\em Annals of Actuarial Science}, 14:150--169.

\bibitem[Tang et~al., 2022]{TWY+22}
Tang, M.-L., Wu, Q., Yang, S., and Tian, G.-L. (2022).
\newblock Dirichlet composition distribution for compositional data with zero
  components: An application to fluorescence in situ hybridization (fish)
  detection of chromosome.
\newblock {\em Biometrical Journal}, 64(4):714--732.

\bibitem[Tsagris and Stewart, 2018]{TS18}
Tsagris, M. and Stewart, C. (2018).
\newblock A dirichlet regression model for compositional data with zeros.
\newblock {\em Lobachevskii Journal of Mathematics}, 39:398--412.

\bibitem[Tsagris and Stewart, 2020]{TS20}
Tsagris, M. and Stewart, C. (2020).
\newblock A folded model for compositional data analysis.
\newblock {\em Australian \& New Zealand Journal of Statistics},
  62(2):249--277.

\bibitem[Tsagris and Stewart, 2022]{TS22}
Tsagris, M. and Stewart, C. (2022).
\newblock A review of flexible transformations for modeling compositional data.
\newblock In He, W., Wang, L., Chen, J., and Lin, C.~D., editors, {\em Advances
  and Innovations in Statistics and Data Science}, pages 225--234. Springer.

\bibitem[Tsagris et~al., 2011]{TPW11}
Tsagris, M.~T., Preston, S., and Wood, A.~T. (2011).
\newblock A data-based power transformation for compositional data.
\newblock In {\em Proceedings of the 4th international workshop on
  Compositional Data Analysis}, Girona, Spain.
\newblock URL: {\url{https://arxiv.org/abs/1106.1451}}.

\bibitem[{World Health Organization}, 1992]{ICD10}
{World Health Organization} (1992).
\newblock The {ICD-10} classification of mental and behavioural disorders.
\newblock URL:
  \url{https://cdn.who.int/media/docs/default-source/classification/other-classifications/9241544228_eng.pdf}.

\bibitem[Zhang et~al., 2023]{ZHH23}
Zhang, X., Huang, F., Hui, F. K.~C., and Haberman, S. (2023).
\newblock {Cause-of-death mortality forecasting using adaptive penalized tensor
  decompositions}.
\newblock {\em Insurance: Mathematics and Economics}, 111:193--213.

\end{thebibliography}

\newpage
\begin{appendices}
\section{Supplementary Information and Results}

\subsection{The Lee-Carter model for modelling  mortality}\label{sec:LC}

This paper applies the LC mortality model after LRA and the $\alpha$-transformation \citep{LC92}. For completeness, this appendix provides a brief review of the LC model for analysis of non-compositional data, that is, RDA. 

Treating causes independently, the LC model fits and predicts central mortality rates by expressing the log mortality rate as a linear function of a time factor with age parameters. For cause $c$, let $m_{t, u, c}$ denote the central death rate for age $u$ in year $t$, which we compute as $m_{t, u, c} = d_{t, u, c}/L_{t, u}$ where the denominator $L_{t, u}$ is the exposure of person-years lived at age $u$. The LC model is then defined as:
\begin{equation}\label{eqn:LC}
    \ln(m_{t, u, c}) = \mu_{u, c} + b_{u, c}k_{t, c} + \epsilon_{t, u, c},
\end{equation}
where $\mu_{u, c}$ represents an age- and cause-specific average mortality over time; $b_{u, c}$ denotes the age- and cause-specific coefficients that vary over time; $k_{t, c}$ denotes a factor of time-varying indices for the level of mortality; and the $\epsilon_{t, u, c}$ denote residual error terms. The model is typically fitted by applying a singular value decomposition to a $U \times T$ matrix the elements of which are given by $\ln(m_{t, u, c})$, after subtracting the average mortality rate over time for a given cause. After fitting, mortality forecasting is performed by modelling the estimated time factors $k_{t, i}$ as an autoregressive integrated moving average time series. The common choice is a simple random walk with drift. We refer the reader to \citet{LC92} for more details regarding parameter estimation of the LC model.

The LC model is commonly used for national forecasts, with its primary advantages including its simplicity, ability to deal with uncertainty, and low requirement for subjective judgement \citep{BK22}. With its simplicity comes a number of limitations, and consequently many variations of LC exist to improve its performance. Among many others, examples include \citet{RH03},  which generalized the LC model to include more than one factor; the \cite{CBD06} model, which is a popular alternative that models the probability of survival rather than the $\log_{10}$ mortality rates; the \cite{LM01} and \cite{BMS02} models, both of which aim to improve the forecasting performance of the LC model \citep{BTS05}; and \citet{LL05} and \citet{GS21}, who apply coherence in the context of mortality modelling, and age-coherent extensions of LC respectively.

\subsection{Additional results for the application to the HCD database} \label{subsec:tuningalpha_additionalresults}

Table~\ref{tab:cv} shows the results from cross-validation for England and Wales's cause of death data. Based on cross-validation, we determined the optimal $\alpha$ value is 0.1 for males and 0.8 for females. This was then applied to produce the results in Section~\ref{sec:cardio_results}.

{Table~\ref{tab:us_cv} similarly shows the results from cross validation for US deaths counts by cause to determine the optimal $\alpha$. Based on cross-validation, we determined the optimal $\alpha$ value is 0.7 for males and 0.9 for females. This was applied to produce the results in Section~\ref{sec:cardio_results_US}.}

\begin{table}[!htb] 
\tabcolsep 0.47in
\caption{Results for validation sets (RMSE and MAE, based on four-fold expanding window cross-validation) to tune $\alpha$, using the $\alpha$-transformation coupled with an LC model for forecasting in our application to England and Wales death counts by cause from \cite{HMD24}, disaggregated for cardiovascular causes of death. Optimal values of $\alpha$ are shown in bold, noting all results are scaled by multiplying by 100.}\label{tab:cv} 
\centering
\begin{tabular}{@{}lcccc@{}} 
\toprule[1.5pt]
& \multicolumn{2}{c}{RMSE} & \multicolumn{2}{c}{MAE}  \\ [0.5ex] $\alpha$ & Male & Female & Male & Female \\
\cmidrule{1-5}
0 (CLR) & 0.1919 & 0.2022 & 0.0985 & 0.0931\\
0 (ILR) & 0.1919 & 0.2022 & 0.0985 & 0.0931\\
0.1     & \bf{0.1992} & 0.2001 & \bf{0.0766} & 0.0727\\
0.2     & 0.2037 & 0.1903 & 0.0791 & 0.0694\\
0.3     & 0.2099 & 0.1813 & 0.0812 & 0.0669\\
0.4     & 0.2542 & 0.1733 & 0.0924 & 0.0649\\
0.5     & 0.2641 & 0.1660 & 0.0961 & 0.0633\\
0.6     & 0.2757 & 0.1595 & 0.1000 & 0.0619\\
0.7     & 0.2882 & 0.1539 & 0.1043 & 0.0607\\
0.8     & 0.3200 & 0.1500 & 0.1135 & \bf{0.0602}\\
0.9     & 0.3347 & \bf{0.1492} & 0.1182 & 0.0613\\
1 (RDA) & 0.3327 & 0.1517 & 0.1174 & 0.0632\\
\bottomrule[1.5pt]
\end{tabular}
\end{table}

\begin{table}[!htb] 
\tabcolsep 0.47in
\caption{Results for validation sets (RMSE and MAE, based on ten-fold expanding window cross-validation) to tune $\alpha$, using the $\alpha$-transformation coupled with an LC model for forecasting in our application to US death counts by cause from \cite{HMD24}, disaggregated for cardiovascular causes of death. Optimal values of $\alpha$ are shown in bold, noting all results are scaled by multiplying by 100.}\label{tab:us_cv} 
\centering
\begin{tabular}{@{}lcccc@{}} 
\toprule[1.5pt]
& \multicolumn{2}{c}{RMSE} & \multicolumn{2}{c}{MAE}  \\ [0.5ex] $\alpha$ & Male & Female & Male & Female \\
\cmidrule{1-5}
0 (CLR) & 0.2320 & 0.3078 & 0.1092 & 0.1195\\
0 (ILR) & 0.2320 & 0.3078 & 0.1092 & 0.1195\\
0.1     & 0.2244 & 0.3101 & 0.0827 & 0.0963\\
0.2     & 0.2146 & 0.2964 & 0.0797 & 0.0929\\
0.3     & 0.2061 & 0.2843 & 0.0771 & 0.0898\\
0.4     & 0.1990 & 0.2736 & 0.0750 & 0.0871\\
0.5     & 0.1933 & 0.2648 & 0.0732 & 0.0848\\
0.6     & 0.1891 & 0.2560 & 0.0717 & 0.0827\\
0.7     & \bf{0.1868} & 0.2485 & \bf{0.0709} & 0.0810\\
0.8     & 0.1871 & 0.2418 & 0.0709 & 0.0794\\
0.9     & 0.1913 & 0.2390 & 0.0721 & \bf{0.0788}\\
1 (RDA) & 0.1998 & \bf{0.2386} & 0.0744 & 0.0791\\
\bottomrule[1.5pt]
\end{tabular}
\end{table}

\subsection{Additional results comparing forecast performance using CLR and ILR transformations with different techniques to replace zero counts}\label{sec:replacementzeros}

\begin{figure}[!htb]
\centering
{\includegraphics[width=15.8cm]{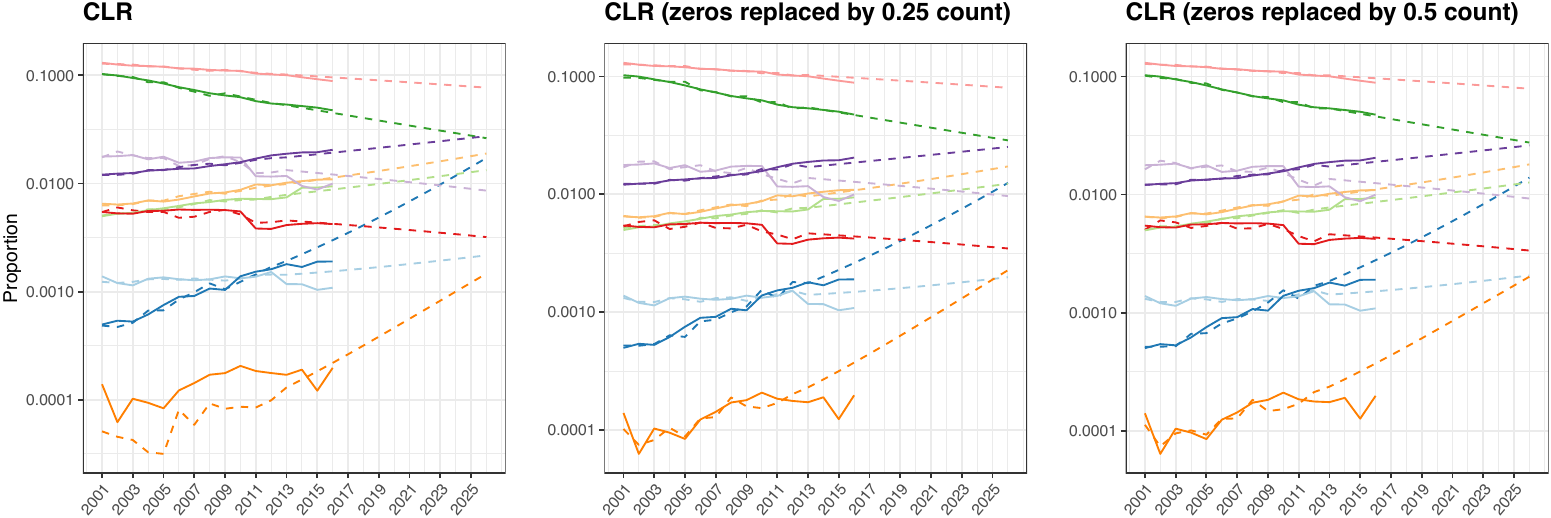}\label{fig:clr_zeros_m}}
\\
{\includegraphics[width=15.8cm]{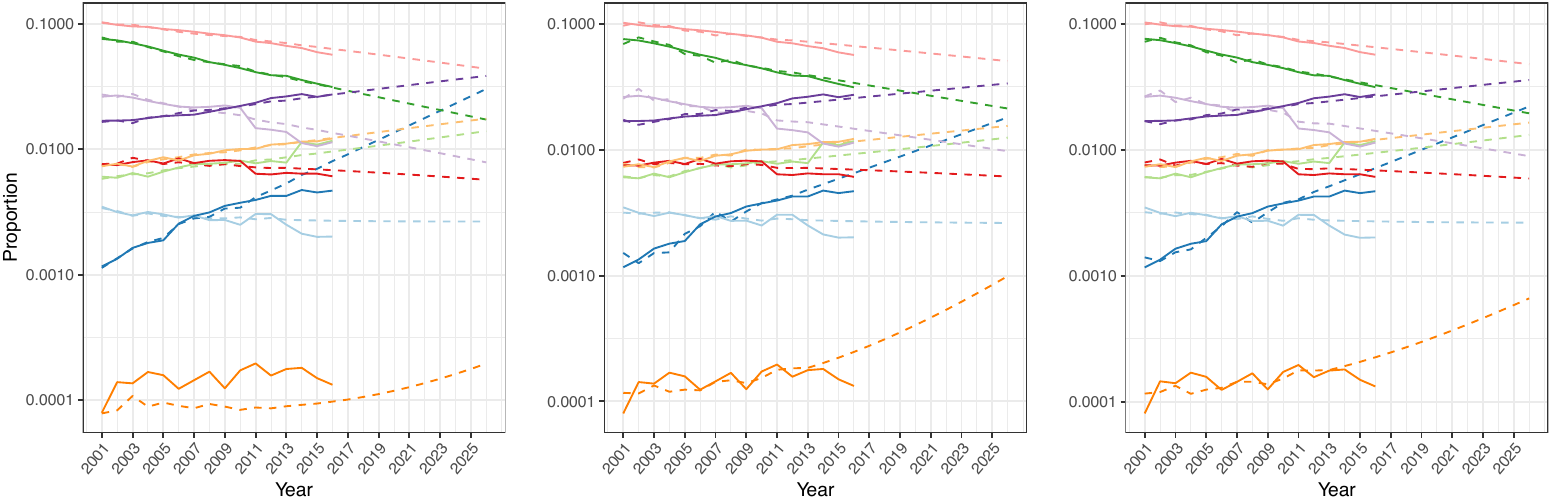}\label{fig:clr_zeros_f}}
\quad
\caption{Forecast of cause-specific mortality up to 2026 in our application to England and Wales death counts by cause from \cite{HMD24}, disaggregated for cardiovascular causes of death. Solid lines represent the observed mortality by cause proportions, and dashed lines show the forecast using the CLR transformation with variations in the treatment of zeros in the data. Mortality by cause is shown for males (top row) and females (bottom row). This figure omits non-cardiovascular causes for presentation purposes.} \label{fig:replacementzeros_clr}
\end{figure}

\begin{figure}[!htbp]
\centering
{\includegraphics[width=15.8cm]{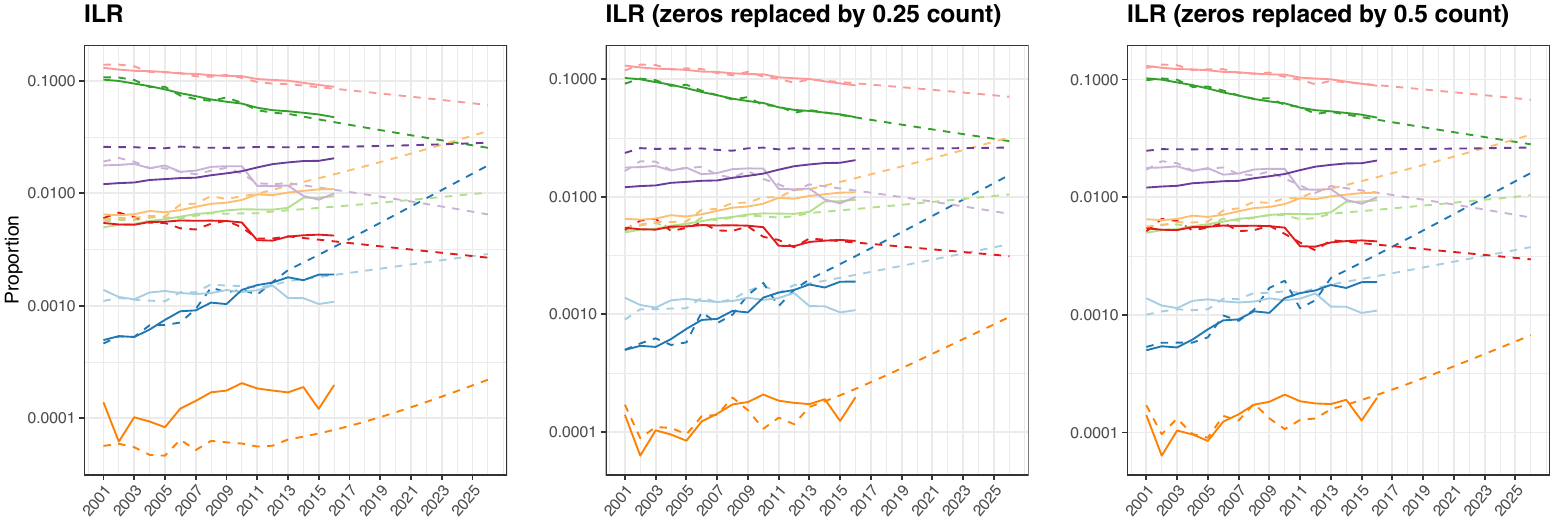}\label{fig:ilr_zeros_m}}
\\
{\includegraphics[width=15.8cm]{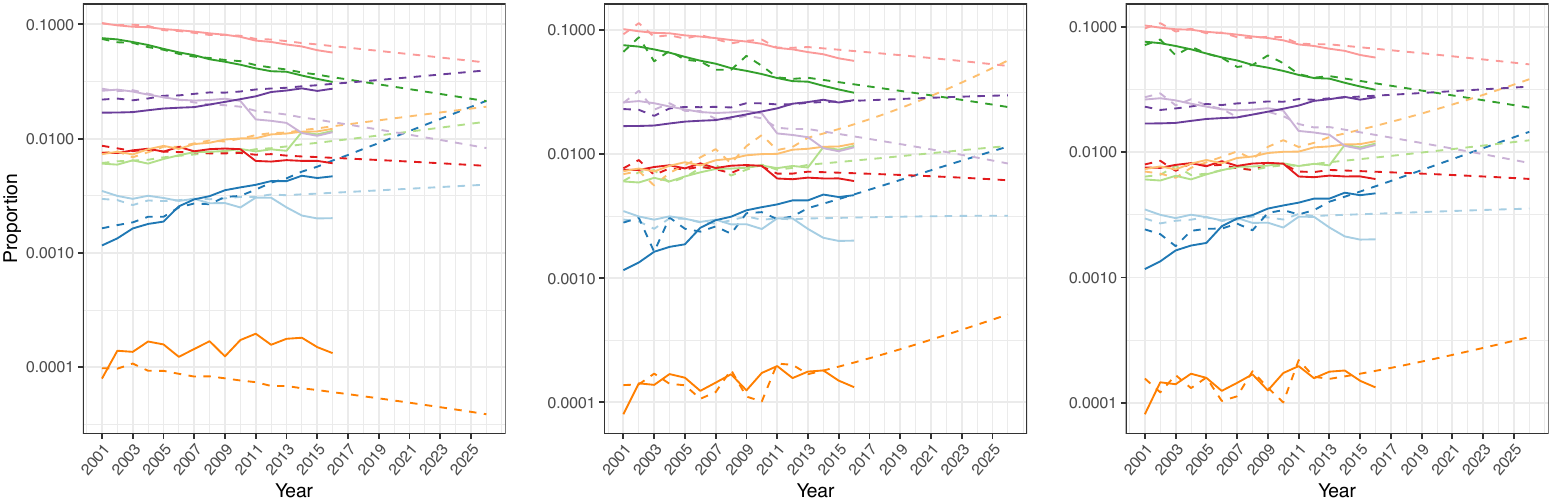}\label{fig:ilr_zeros_f}}
\quad
\caption{Forecast of cause-specific mortality up to 2026 in our application to England and Wales death counts by cause from \cite{HMD24}, disaggregated for cardiovascular causes of death. Solid lines represent the observed mortality by cause proportions, and dashed lines show the forecast using the ILR transformation with variations in the treatment of zeros in the data. Mortality by cause is shown for males (top row) and females (bottom row). This figure omits non-cardiovascular causes for presentation purposes.} \label{fig:replacementzeros_ilr}
\end{figure}

We further compared the performance of CLR and ILR forecasts when zero counts are replaced by 0.25 or 0.5 for both England and Wales and US death counts by cause of death. This was applied for both male and female death counts on both sets of data for completeness. These results are included in Sections~\ref{sec:cardio_results} and~\ref{sec:cardio_results_US}. For England and Wales, Figures~\ref{fig:replacementzeros_clr} and~\ref{fig:replacementzeros_ilr} show the visualisations of the forecasts when different zero replacement approaches are used. The forecast and trends change and are sensitive to the method of zero replacement. The $\alpha$-transformation presents a statistical approach that removes this sensitivity.

\begin{figure}[!htb]
{\includegraphics[width=17.5cm]{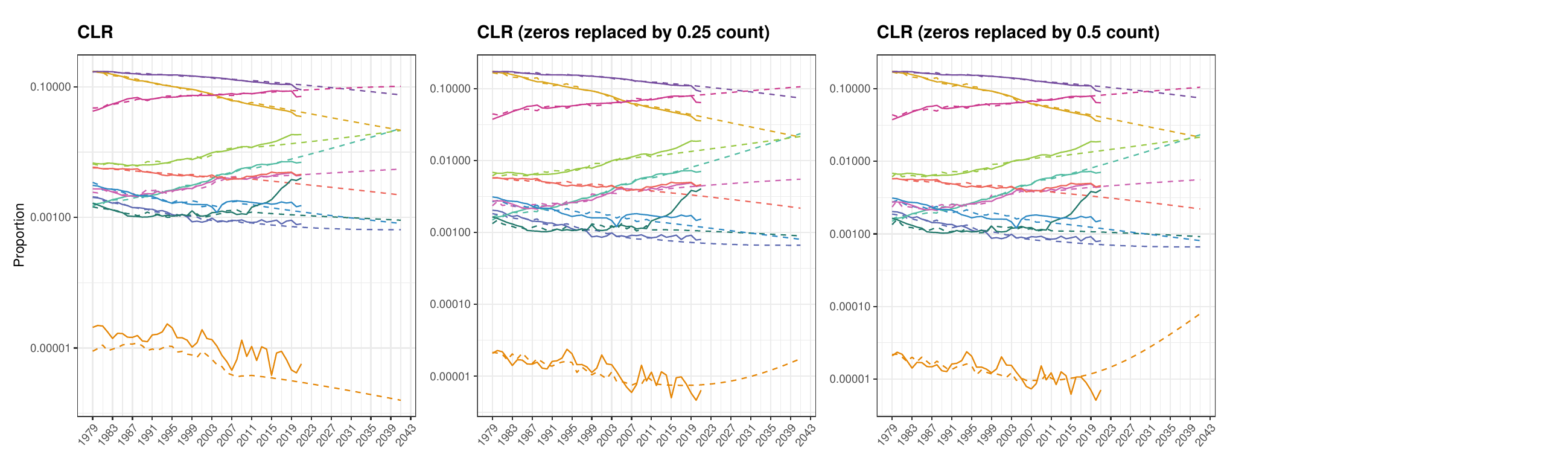}\label{fig:us_clr_zeros_m}}
\\
{\includegraphics[width=16.2cm, height=5cm]{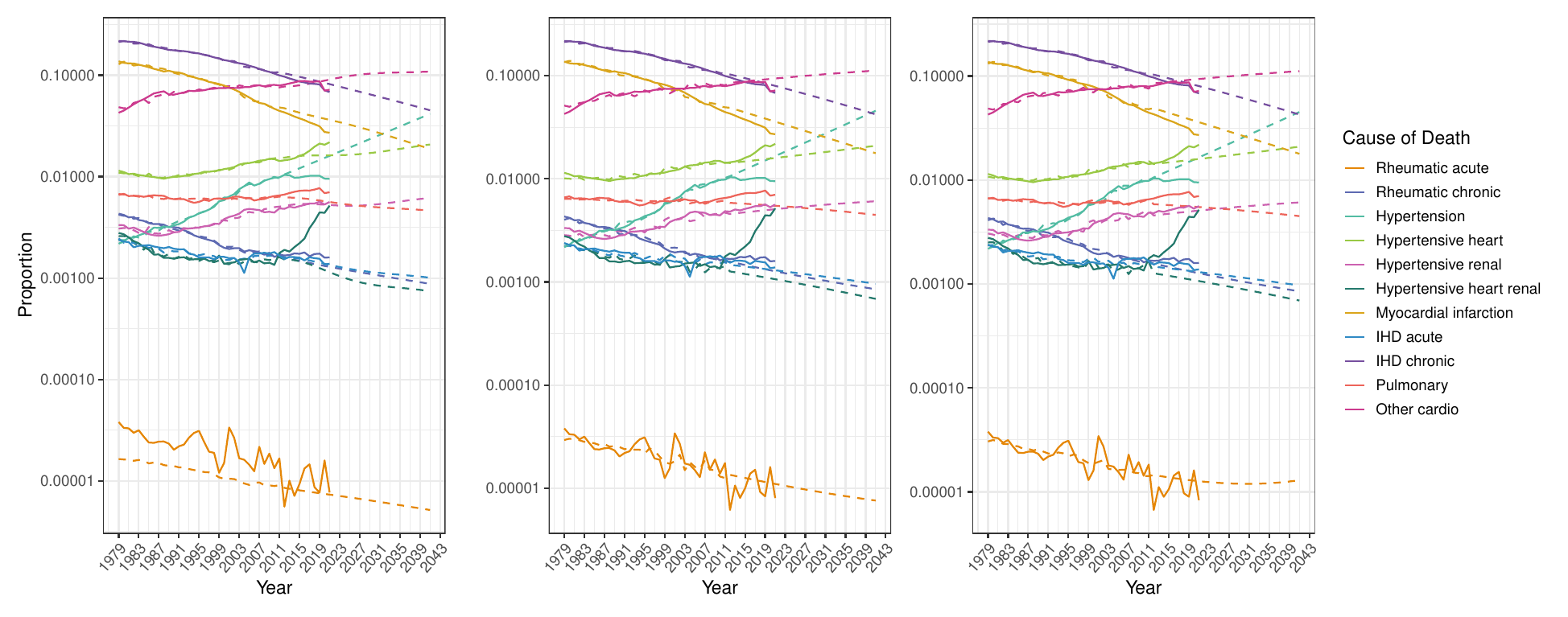}\label{fig:us_clr_zeros_f}}
\quad
\caption{Forecast of cause-specific mortality up to 2051 in our application to US death counts by cause from \cite{HMD24}, disaggregated for cardiovascular causes of death. Solid lines represent the observed mortality by cause proportions, and dashed lines show the forecast using the CLR transformation with variations in the treatment of zeros in the data. Mortality by cause is shown for males (top row) and females (bottom row). This figure omits non-cardiovascular causes for presentation purposes.} \label{fig:us_replacementzeros_clr}
\end{figure}

\begin{figure}[!htbp]
{\includegraphics[width=17.5cm]{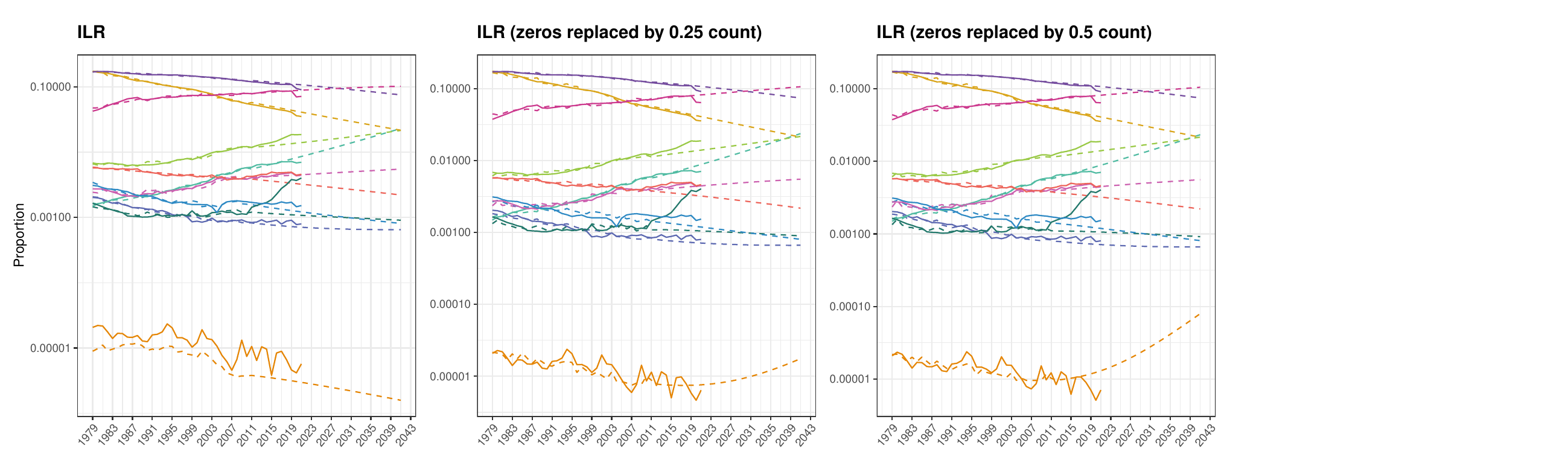}\label{fig:us_ilr_zeros_m}}
\\
{\includegraphics[width=16.2cm, height=5cm]{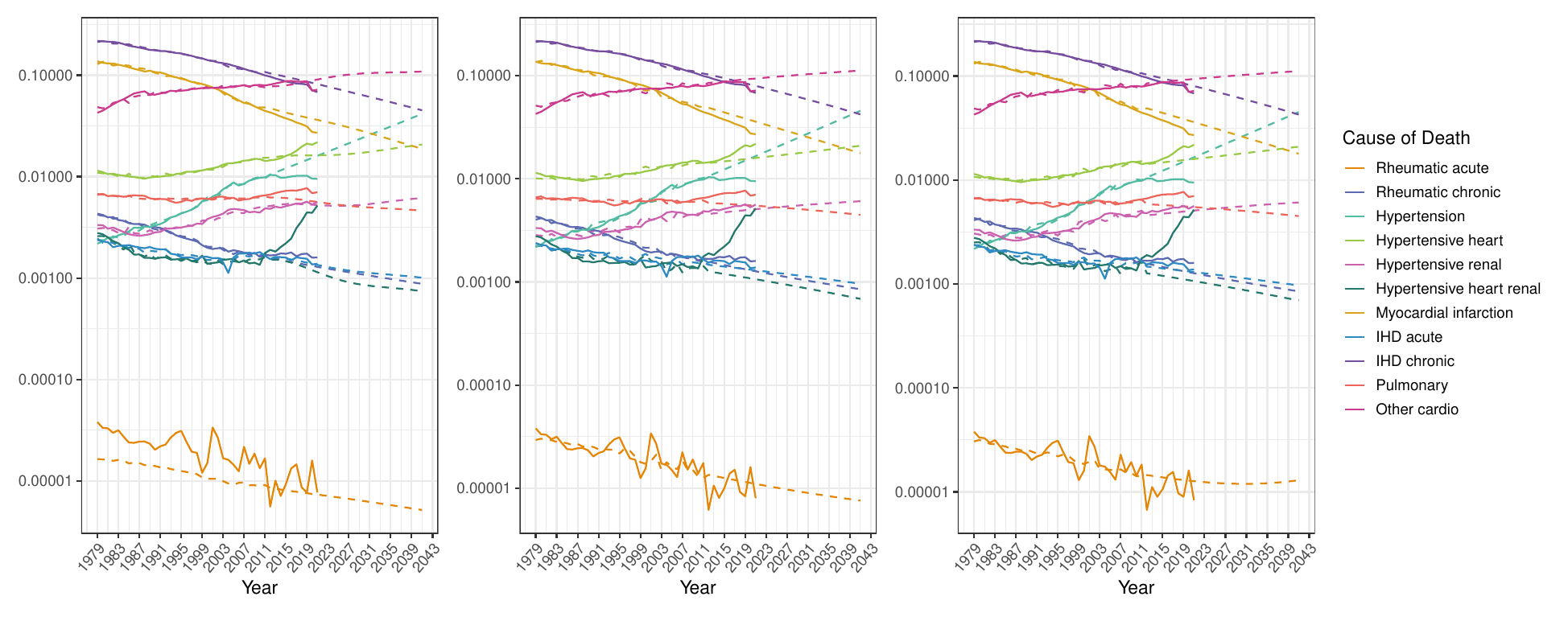}\label{fig:us_ilr_zeros_f}}
\quad
\caption{Forecast of cause-specific mortality up to 2051 in our application to US death counts by cause from \cite{HMD24}, disaggregated for cardiovascular causes of death. Solid lines represent the observed mortality by cause proportions, and dashed lines show the forecast using the ILR transformation with variations in the treatment of zeros in the data. Mortality by cause is shown for males (top row) and females (bottom row). This figure omits non-cardiovascular causes for presentation purposes.} \label{fig:us_replacementzeros_ilr}
\end{figure}

Similarly, for US death counts, Figures~\ref{fig:us_replacementzeros_clr} and~\ref{fig:us_replacementzeros_ilr} show visualisations of the forecasts when different zero replacement approaches are used. It is worth noting that longer term trends are still impacted by different approaches to replace zeros, despite the US data set having a longer history compared to the England and Wales death counts by cause.

\FloatBarrier

\subsection{Sensitivity analysis of the choice of \texorpdfstring{$\alpha$}{alpha}}\label{app:sensitivity_alpha}

A sensible question to ask then is how sensitive were the results to the particular chosen values of $\alpha$ as long as we were within this tolerance range. Based on additional testing, we found that results remain largely unaffected when $\alpha$ was specified within 0.1. 

The optimal $\alpha$ for England and Wales death counts (Section~\ref{sec:cardio_results}) is 0.1 for males, resulting in RMSE and MAE of 0.1818 and 0.1046 respectively. For $\alpha = 0.09$, the resulting RMSE and MAE is 0.1832 and 0.1055. For $\alpha = 0.11$, the resulting RMSE and MAE is 0.1806 and 0.1037. Here, the results improve when $\alpha = 0.11$, compared to specifying $\alpha$ to the nearest 0.1. However, the resulting inferences around mortality forecasts by cause are unchanged.

We perform a similar exercise on the optimal alphas for US data, where there is a longer history of death counts. For example, the optimal $\alpha$ for US females (Section~\ref{sec:cardio_results_US}) is 0.9, resulting in RMSE and MAE of 0.2516 and 0.1238, respectively. For $\alpha = 0.91$, the resulting RMSE and MAE are 0.2528 and 0.1243. For $\alpha = 0.89$, the resulting RMSE and MAE are 0.2504 and 0.1233, an improvement to the selected optimal $\alpha = 0.90$. Moreover, the resulting inferences from the forecast were largely unchanged in terms of shape and trend in the forecast of cause-specific mortality. 
\end{appendices}

\end{document}